\documentclass[iop]{emulateapj}
\usepackage{apjfonts}
\usepackage{graphicx}
\usepackage[usenames]{color}
\usepackage{amsmath}
\newcommand       \ISO          {{\it ISO}}
\newcommand       \Spitzer      {{\it Spitzer}}

\newcommand       \JWST       {{\it JWST}}
\newcommand       \IRS          {{\rm IRS}}

\newcommand       \MIPS         {{\rm MIPS}}

\newcommand       \LongISO      {{\it Infrared Space Observatory}}
\newcommand       \LongSpitzer  {{\it Spitzer Space Telescope}}
\newcommand       \LongHerschel {{\it Herschel Space Observatory}}

\newcommand       \LongJWST   {{\it James Webb Space Telescope}}
\newcommand       \LongIRS      {{\rm Infrared Spectrograph}}

\newcommand       \Angstrom     {\,\textnormal{\AA}}
\newcommand       \um           {\,{\micron}}
\newcommand       \mic          {\hbox{$\um$}}

\newcommand       \K            {\,{\rm K}}

\newcommand       \LSun         {\,\ensuremath{L_\sun}}
\newcommand       \Lo           {\,\ensuremath{L_\sun}}
\newcommand       \MSun         {\,\ensuremath{M_\sun}}
\newcommand       \Mo           {\,\ensuremath{M_\sun}}
\newcommand       \LIR          {\ensuremath{L_{\rm IR}}}
\newcommand       \Lbol         {\ensuremath{L_{\rm bol}}}
\newcommand       \tV           {\mbox{$\tau_{\rm V}$}}
\newcommand       \Sil          {\mbox{$S_{\rm sil}$}}
\newcommand       \Is           {\mbox{$I_{\rm rep}$}}
\newcommand       \Tin          {\mbox{$T_{\rm in}$}}

\newcommand{\E}[1]  {\hbox{$10^{#1}$}}
\newcommand       \about        {\hbox{$\sim$}}
\newcommand       \x            {\hbox{$\times$}}
\newcommand       \xAGN     {\hbox{$\textrm{x}_\textrm{AGN}$}}

\newcommand       \xShell    {\hbox{$\textrm{x}_\textrm{shell}$}}
\newcommand       \xPAH    {\hbox{$\textrm{x}_\textrm{PAH}$}}

\def\rem#1{\relax}

\shorttitle{The Nature of Deeply Buried ULIRGs}
\shortauthors{Marshall et al.}
\begin{document}
\title{The Nature of Deeply Buried Ultraluminous Infrared Galaxies: \\
A Unified Model for Highly Obscured Dusty Galaxy Emission}

\author{
  J.~A. Marshall,\altaffilmark{1}
  M. Elitzur,\altaffilmark{3,4}
  L. Armus,\altaffilmark{2}
  T. Diaz-Santos,\altaffilmark{5}
  and V. Charmandaris\altaffilmark{6,7}
}

\altaffiltext{1}{Glendale College, Glendale, CA 91208; jmarshall@glendale.edu}

\altaffiltext{2}{Spitzer Science Center, California Institute of Technology,
Pasadena, CA 91125}

\altaffiltext{3}{Department of Physics and Astronomy, University of Kentucky,
Lexington, KY 40506-0055}

\altaffiltext{4}{Astronomy Department, University of California, Berkeley, CA
94720-3411}

\altaffiltext{5}{Universidad Diego Portales, Calle Vergara 201, Santiago,
Chile}

\altaffiltext{6}{Department of Physics, University of Crete, GR-71003,
Heraklion, Greece}

\altaffiltext{7}{Institute for Astronomy, Astrophysics, Space Applications \&
Remote Sensing, National Observatory of Athens, GR-15236, Athens, Greece}


\begin{abstract}

We present models of deeply buried ultraluminous infrared galaxy (ULIRG) spectral energy
distributions (SEDs) and use them to construct a three-dimensional diagram for diagnosing the nature
of observed ULIRGs. Our goal is to construct a suite of SEDs for a very simple model ULIRG 
structure, and to explore how well this simple model can (by itself) explain the full range of observed ULIRG
properties. We use our diagnostic to analyze archival \LongSpitzer\ \IRS\ spectra of ULIRGs and find that: 
(1) In general, our model does provide a comprehensive explanation of the distribution of mid-IR ULIRG properties; 
(2) $>$75\% (in some cases 100\%) of the bolometric luminosities of the most deeply buried ULIRGs {\it must} be 
powered by a dust-enshrouded active galactic nucleus; 
(3) an unobscured ``keyhole'' view through $\la$10\% of the obscuring medium surrounding a deeply buried
ULIRG is sufficient to make it appear nearly unobscured in the mid-IR; and
(4) the observed absence of deeply buried ULIRGs with large PAH equivalent widths is naturally explained 
by our models showing that deep absorption features are ``filled-in'' by small quantities of foreground unobscured 
PAH emission (e.g., from the host galaxy disk) at the level of $\sim$1\% the bolometric nuclear luminosity. 
The modeling and analysis we present will also serve as a powerful tool for interpreting the high angular 
resolution spectra of high-redshift sources to be obtained with the \LongJWST.

\end{abstract}

\keywords{
  galaxies: active ---
  galaxies: ISM ---
  galaxies: starburst ---
  infrared: galaxies}

\section{Introduction}
\label{sec:Introduction}

With $\LIR \equiv L_{8-1000\um} > 10^{12} \LSun$, Ultraluminous Infrared Galaxies (ULIRGs) 
emit more in the infrared than typical galaxies emit at all wavelengths combined. This implies that ULIRGs 
contain a powerful, often unresolved, central source covered by $\ga10^8-10^9 \MSun$ of dust 
\citep{Marshall07, 2010A&A...523A..78D} that captures the bulk of the outgoing radiation from the central source(s) 
and reprocesses it into the infrared. The exact origin of the central emission in ULIRGs has been the subject 
of debate for years and can plausibly be attributed only to accretion onto a supermassive black hole 
(i.e., an AGN) and/or powerful bursts of star formation. Since the central radiation is mostly (or completely) 
obscured by the surrounding dust, unprocessed X-ray or UV photons providing \textit{direct} evidence about 
the nature of the central engine are generally unavailable. But since such infrared-luminous galaxies are believed 
to be the largest contributors to the far-IR background and star-formation energy density at $z \approx 1$--$3$ 
\citep{Blain02, Elbaz03, Murphy2011, Magnelli2013}, an accurate diagnosis of the mechanism by which their 
power is generated is critical for obtaining an accurate picture of the star-formation and galaxy evolution history 
of the universe. Thus, astronomers have shifted their attention to searching the \textit{indirect} evidence about 
the nature of the obscured central source that can be gleaned from reprocessed infrared radiation.

Initial progress towards this goal was made with the \LongISO\ (\ISO). Diagnostic diagrams based upon the 
mid-IR continuum slope, AGN-indicating high-ionization mid-IR emission lines (e.g. [\ion{O}{4}] and [\ion{Ne}{5}]),
and the strengths of the star-formation-indicating $6.2$ and $7.7\um$ polycyclic aromatic hydrocarbon (PAH) 
emission features were constructed to quantify the relative contributions from accretion and star-formation to the
bolometric luminosities of a small number of local ULIRGs \citep[e.g.][]{Genzel98, Lutz98, Laurent00, Tran01}. 
These observations provided evidence that local ULIRGs are powered primarily by star-formation, but with the 
fraction of AGN activity increasing with bolometric luminosity.

\begin{figure*}
  \epsscale{1.18}
  \plotone{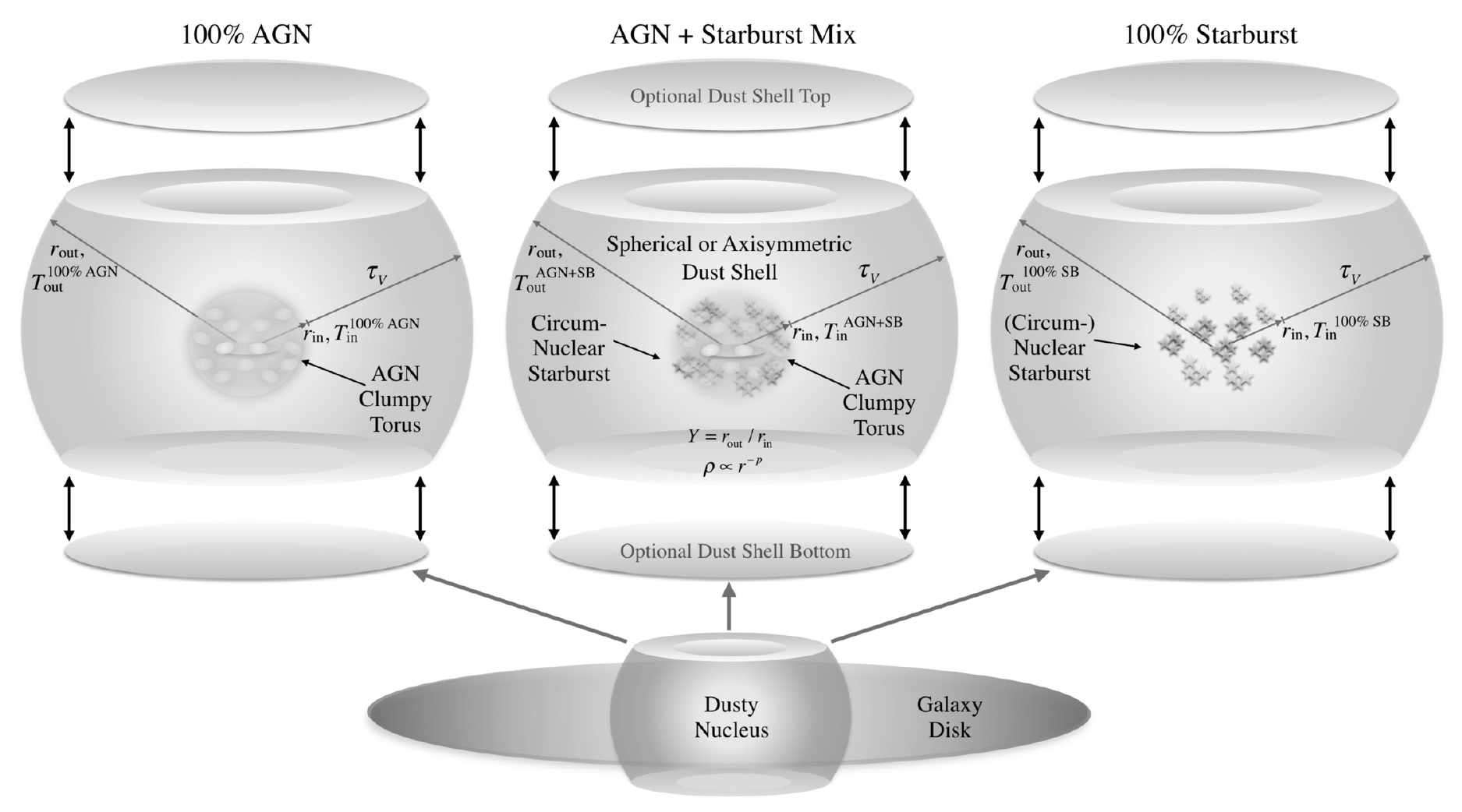}
 
\caption{As shown in this schematic view, our simple ULIRG models are made up of two primary components: 
(1) a dusty nucleus powered by AGN activity (\textit{left}), star-formation (\textit{right}), or a combination of 
both (\textit{center}); and (2) a galactic disk. The SED of the obscured nucleus is obtained by modeling the 
radiative transfer of emission from the nuclear power source(s) through a geometrically and optically thick 
spherical or axisymmetric dust shell. The disk component is dominated in the mid-IR by 
relatively unobscured PAH emission. Note that the dust shell ``tops'' and ``bottoms'' are labeled 
``optional'' since the radiative transfer solutions are (approximately) valid for both spherical and axisymmetric 
geometries over the infrared wavelengths used in this work.}
  \label{fig:ULIRGModel}
  \epsscale{1.0}
\end{figure*}

With its expanded mid-IR coverage and much higher sensitivity, the \LongIRS\ \citep[\IRS;][]{Houck04} on the 
\LongSpitzer\ \citep{Werner04} has provided an abundance of additional data, including mid-IR spectra of large
samples of starburst galaxies \citep{Brandl06}, AGNs \citep{Hao05, Weedman05, Schweitzer06, HerCab2015}, 
local LIRGs \citep{Armus09, Petric11, Stierwalt2013, Stierwalt2014}, 
ULIRGs \citep{Armus06, Armus07, Desai07, Nardini08}, and their high-redshift analogs
\citep[e.g.][]{Houck05, Yan05, Lutz05, Sajina07, Pope08, Huang09, Kirkpatrick2012}.
More recently, the expanded far-IR coverage of the \LongHerschel\ \citep{Pilbratt2010} has opened 
a new window into the far-IR universe and improved our understanding of the cooling of the 
interstellar medium (ISM) within dusty star-forming galaxies 
\citep[e.g.,][]{Rosenberg2015, DiazSan2013, DiazSan2014}.

While this wealth of additional data has significantly improved our understanding of dust-enshrouded 
galaxies \citep[e.g.,][]{Levenson07, Spoon07, DiazSan2010, DiazSan2011, Stierwalt2014}, the nature of the class 
of deeply buried ULIRGs---those whose nuclei are so enshrouded that most or all 
characteristic AGN emission is obscured---remains elusive. Of particular interest, some of the most  
obscured ULIRGs with the most extreme silicate absorption features also display, somewhat paradoxically, 
relatively flat (i.e.,``warm'') mid-infrared slopes. Understanding such an extreme mid-IR spectrum, seen for the 
first time in IRAS\,08572+3915 \citep{Armus07}, has been the subject of detailed radiative transfer 
modelling by \citet{2014MNRAS.437L..16E}, which revealed that reproducing the spectrum requires an edge-on 
inclination and the use of tapered disk models. 

The notorious complexity of mid-IR emission from dusty galaxies is both a blessing and a curse: the wealth of 
features provide numerous possible diagnostic signals, but disentangling individual signals from one another to 
discern their relationship to the nature of the underlying power source is challenging. In particular, 
emission from silicate and carbonaceous dust grains heated to different characteristic temperatures, stellar 
photospheres, PAHs, and atomic and molecular lines---all of which may pass through and be reprocessed by 
different amounts of obscuring dust---combine to create a diverse variety of ULIRG spectral energy distributions 
(SEDs). This inherent richness of information, along with the aforementioned difficulty of diagnosing the 
nature of deeply buried ULIRGs via methods based upon single measurable quantities (such as PAH feature 
equivalent width or [\ion{Ne}{5}] emission line strength), suggests the possibility of a more effective approach 
based upon the comparison of multiple mid-IR spectral properties to those obtained from a suite of radiative
transfer models---such as those presented in this work. 

As an added benefit, these models can also provide a powerful tool for exploring the basic geometries of highly 
obscured nuclei, and should therefore prove to be of great use for analyzing the spectroscopic observations 
at smaller physical size-scales that will soon be obtained using next-generation observatories such as the 
\LongJWST\ (\JWST).

This paper presents our approach for producing model SEDs of deeply buried ULIRGs and provides 
quantitative tools for investigating the nature of ULIRG nuclei. In \S\ref{sec:Modeling}, 
we use the radiative transfer code DUSTY \citep{DUSTY} and the observed spectra of low-obscuration starburst 
galaxies and AGN to construct models of highly obscured ULIRG SEDs that are powered by star-formation, 
AGN activity, or a combination of both. In \S\ref{sec:SpectralIndicators}, we discuss spectral diagnostics and calculate the 
mid-IR slope, PAH equivalent width, and silicate emission or absorption feature strength for each of our models 
and for a sample of \Spitzer\ \IRS\ spectra of AGNs, starburst galaxies, and ULIRGs. Finally, in 
\S\S\ref{sec:diagnostic} and \ref{sec:discussion}, we use these values to construct a three-dimensional diagram for 
diagnosing the nature of deeply buried ULIRG nuclei, and we use this diagnostic to analyze our sample of ULIRGs.

\section{Modeling ULIRG SED\MakeLowercase{s}}
\label{sec:Modeling}

We have developed a method for calculating infrared SEDs of a very simple 
dust-enshrouded ULIRG structure. As shown in Figure~\ref{fig:ULIRGModel}, such a system is 
assumed to be comprised of (1) a central power source buried within an enshrouding shell of dust, and (2) 
a host galaxy disk. The nuclear component of our model is powered by emission from an AGN, a burst of star 
formation, or some combination thereof. If the obscuring shell surrounding the nuclear power source(s) is 
optically thick, spherically symmetric, and smoothly distributed (i.e., not clumpy), \textit{direct} emission 
from the central power source(s) cannot be seen along any line of sight outside the shell. 
If the surrounding dust shell is instead clumpy, or if it is axisymmetric and 
viewed from a favorable angle, some fraction of the emission from the central power source may escape 
unimpeded even for lines of sight in which the nuclear power source(s) are generally obscured.

Our radiative transfer solutions and the SEDs derived from them hold for both spherical and 
(in the situations described below) axial symmetry. This has been discussed previously by 
\citet{1997A&A...318..879M}, where it was shown that the dust temperature distribution in a sphere provides 
a good approximation to the solution of something as non-spherical as a thick flared disk. Thus, to good 
approximation, our models are applicable so long as an observer does not have a direct view of the hot inner 
edge of the obscuring shell (i.e., they must be looking ``through'' the spherical or axisymmetric dust shell). We note 
that these simplifying assumptions made in constructing our models mean that detailed properties 
of some complex axisymmetric or completely asymmetric systems---possibly including the deeply 
obscured ULIRGs IRAS\,08572+3915 \citep{Armus07, 2013ApJ...775L..15R} and 
Arp\,220 \citep[e.g.,][]{2017ApJ...836...66S}---may not be well-represented by our results. 
However, as demonstrated in this work, we emphasize that despite these limitations, our models provide a 
powerful tool for exploring the physical properties of the majority of the population of ULIRGs for which 
adequate data exist.

The following sections provide a detailed description of the assumptions and parameterizations made
to model emission from the obscured nucleus (\S\S\ref{sec:ModelingCentralEngine}, \ref{sec:ModelingShellEmission}, 
and \ref{sec:ModelingNuclearEmission}); the host galaxy disk (\S\ref{sec:ModelingDiskEmission}); and ultimately 
the entire ULIRG (\S\ref{sec:ModelingTotalEmission}).

\subsection{The ULIRG Central Engine}
\label{sec:ModelingCentralEngine}

Buried beneath the obscuring shell of dust at the heart of a ULIRG lies its central AGN and/or starburst power source.
\citet{Sirocky08} found that as long as the emission from this central source is concentrated in the optical/UV, its spectral 
shape has little effect on the spectrum of emission emerging from a surrounding dust shell. In particular, their 
Figure~1 shows that the SEDs emitted by dusty obscuring shells with radial $\tV \ge 10$ are nearly identical for central 
heating by an A0 star, an AGN, or the optical/UV light from a starburst galaxy. But even if the spectrum of emission from the
dust in such a shell is independent of the heating source, the total spectrum obtained by combining the emission from the 
obscuring dust and the obscured central source is not. The total emission includes light from the dusty obscuring shell as 
well as the obscured optical, UV, and possibly infrared emission from the various components of the extended central 
heating source---i.e., the clumpy dusty torus immediately surrounding the AGN accretion disk and/or the dusty knots 
surrounding (or in close proximity to) newly formed stars. A realistic model of the mid-IR properties of deeply obscured 
ULIRGs therefore must begin with a realistic model of the emission spectrum responsible for heating the obscuring material.

Since our goal in this paper is not to produce a ``first-principles'' model of ULIRGs 
\citep[for examples of such comprehensive ``first-principle'' approaches see, e.g.,][]{2007A&A...461..445S,
2009MNRAS.399..615R, 2014MNRAS.437L..16E, 2015A&A...583A.120S}, but is to construct a very 
simple model that reproduces the spectral properties of as many sources as possible, we choose a phenomenological 
approach and model the central engine heating sources using linear combinations of a fitted prototype AGN SED and
either a low-obscuration or high-obscuration starburst nucleus SED. The quasar PG\,0804+761 serves as our prototype 
AGN, and NGC\,7714 and NGC\,2623 serve as our prototype low-obscuration and high-obscuration starburst nuclei, 
respectively. These sources were selected as representative prototypes after carrying out a careful examination of 
nearly $\sim$500 high-quality \Spitzer\ \IRS\ spectra of local galaxies analyzed by our team as part of \Spitzer\ GTO 
and GO programs.

\begin{figure*}
  \epsscale{0.75}
  \plotone{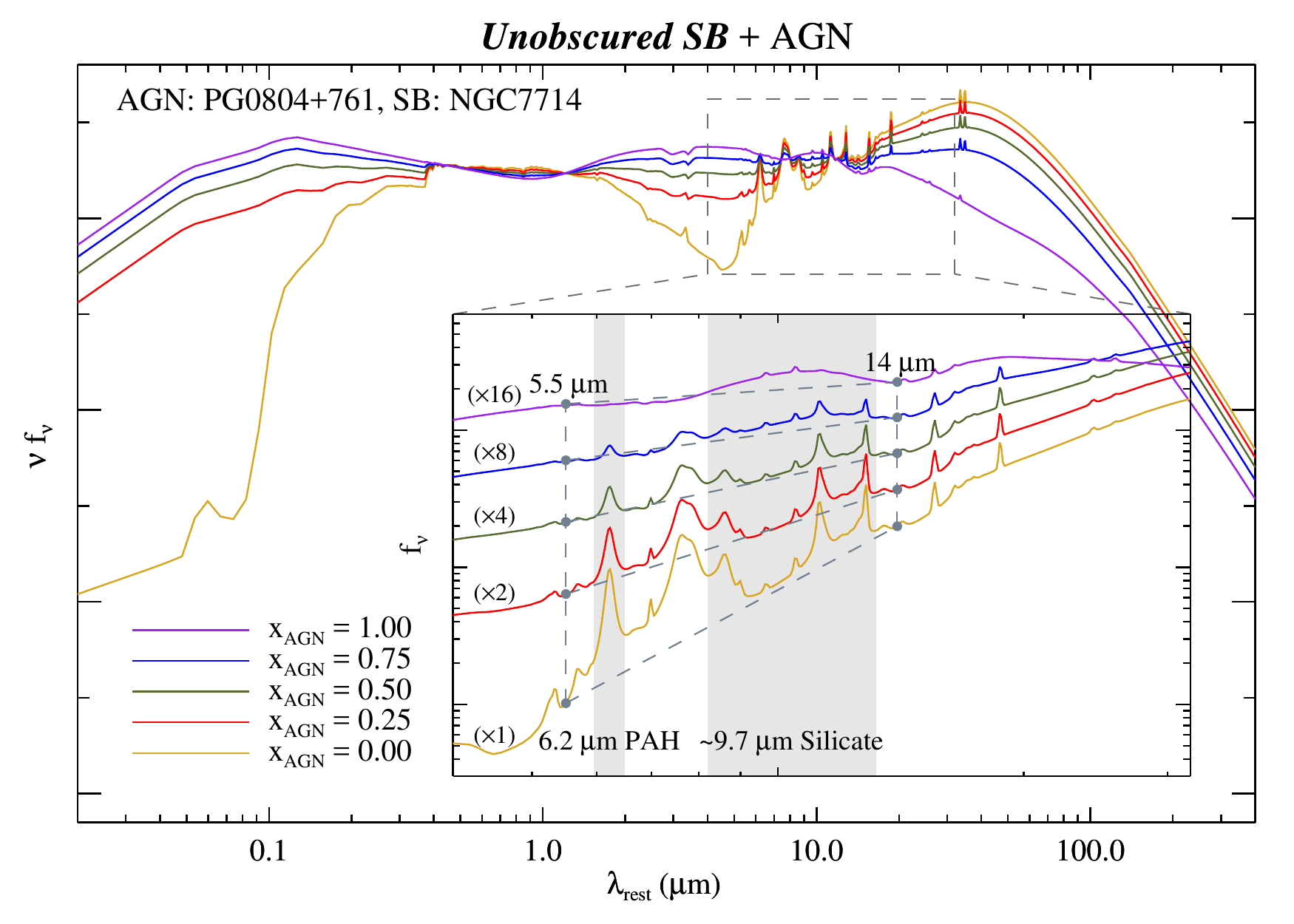}
  \plotone{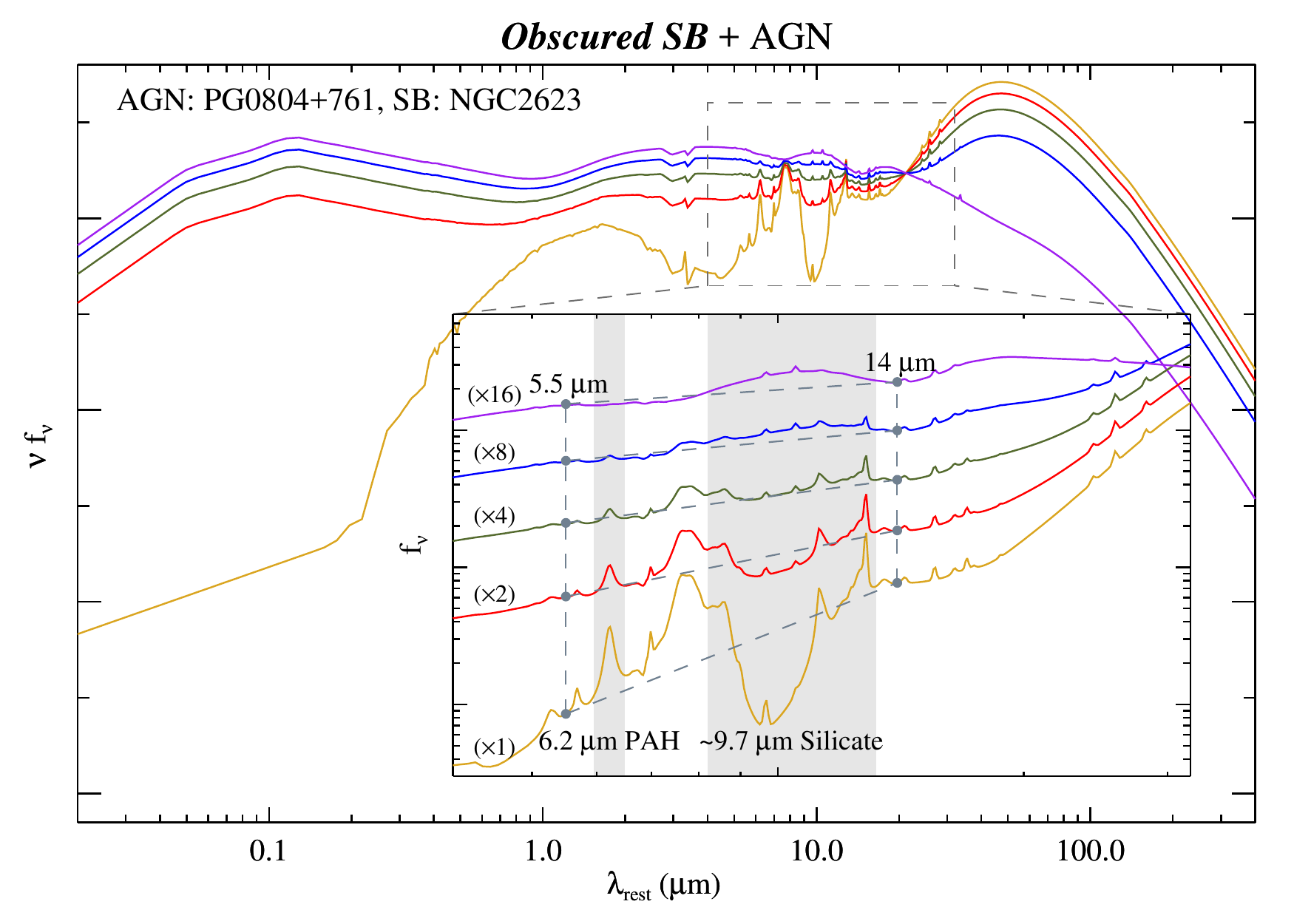}
 
\caption{UV-to-far-IR and 4--32\um\ mid-IR (\textit{inset}) SEDs of our model central engine heating sources. 
Each panel shows linear combinations of a low-obscuration AGN (modeled with the nuclear SED of PG\,0804+761) 
and either a relatively unobscured or obscured starburst (SB) nucleus (modeled with the nuclear SEDs of NGC\,7714 and 
NGC\,2623, respectively). The fraction of the bolometric luminosity from the AGN component is given by \xAGN;
where $\xAGN = 0$, 0.25, 0.5, 0.75, and 1 indicate AGN contributions of 0\%, 25\%, 50\%, 75\%, and 100\%, respectively. 
Flux densities for the full SEDs are scaled such that $\int f_\nu d\nu = 1$, and mid-IR SEDs are further scaled 
(for clarity) by the factors indicated. The spectral ranges of the 6.2\mic\ PAH and 9.7\mic\ silicate features are highlighted, 
and the pairs of dots and dashed lines illustrate the mid-IR continuum slope used in our analysis 
(see \S\S\ref{sec:SpectralSlope} and \ref{sec:AnalysisModelsData}).}
  \label{fig:InputSEDs}
  \epsscale{1.0}
\end{figure*}

Figure~\ref{fig:InputSEDs} shows the central engine SEDs produced by this linear combination 
procedure for both the relatively unobscured and more heavily obscured starburst nuclei heating sources (left and right 
panels, respectively). For each pair of heating sources, we vary the relative contribution from the AGN and 
starburst---parametrized by $\xAGN$, where $\xAGN$ varies from 0 to 1 for AGN contributions between 
0\% and 100\%---to simulate the full range of potential central engines heating sources.

The fitted SEDs of our prototype AGN and starburst galaxies were derived from the ``nuclear'' fits from
\citet{Marshall07} in which the far-IR is constrained by \Spitzer\ \MIPS\ photometry extracted over spatial scales that 
match the \Spitzer\ \IRS\ slit widths as closely as possible to minimize the amount of included host galaxy disk emission. 
The SEDs of both template starburst galaxies were constructed by subtracting the fitted $T \approx 30\,\textrm{K}$ ``cold'' 
dust component from the total fit to further remove any emission from the host galaxy disk. This choice is supported by the 
detailed analysis presented in that paper where it is shown that the fitted quantity of extinguished optical/UV light for both 
starburst sources is sufficient to power all but the ``cold'' dust component. No such procedure was required for the  
template AGN since its ``nuclear''  and ``global'' fits in \citet{Marshall07} are essentially identical---indicating very little 
contribution from a host galaxy disk to its total emission.

\subsection{The Obscuring Dust Shell}
\label{sec:ModelingShellEmission}

We use the radiative transfer code DUSTY\footnote{http://faculty.washington.edu/ivezic/dusty\_web/} to calculate
the SED of emission emerging from a spherical dust shell surrounding the AGN and/or starburst central heating source(s) of
\S\ref{sec:ModelingCentralEngine}. DUSTY \citep{IE97} calculates the emergent SED after the input radiation is processed and 
altered via scattering, absorption, and re-emission as it passes through the obscuring dust. Taking full advantage of the 
scaling properties of the radiative transfer problem, DUSTY requires the minimal set of input parameters for 
thermal emission from radiatively heated dust. In particular, the only relevant property of the central source is the spectral 
shape of its radiation---its luminosity is irrelevant \citep{IE97}.

For the obscuring dust we adopt standard properties of interstellar grains. The grain-size distribution function is from \citet{MRN}, 
although we note that our results are not sensitive to this choice since, as stated in \citet{Sirocky08}, changes in this 
distribution function significantly affect the emergent SED only at near-IR and shorter wavelengths, and these are shorter 
than the wavelengths of interest in our analysis. We adopt a 53\% silicate and 47\% graphite grain composition, and use 
the optical properties of silicate and graphite grains from \citet{Ossenkopf92} and \citet{Draine03}, respectively. These 
choices were adopted from the work of \citet{Sirocky08} who find that this composition and distribution function 
best reproduces the observed ratio of the $9.7$ to $18\um$ silicate absorption feature strengths in ULIRGs.

Unlike the models of \citet{2007A&A...461..445S} and \citet{2009MNRAS.399..615R}, the emission 
from our obscuring dust shell is devoid of PAH emission (since DUSTY itself does not model stochastic heating of small 
grains). In light of the fact that we use PAH feature equivalent width as a diagnostic in our analysis, this would appear to 
be a significant omission. However, as described in detail in \S\ref{sec:PAH}, for the geometrical assumptions of our model, 
the total PAH emission from a galaxy will be dominated by the PAH emission from our nuclear heating source SEDs, 
and the obscuring dust shell will not contribute significantly to the total PAH output. In brief, this is because
PAH molecules must be bathed in a sufficiently dilute yet sufficiently hard radiation field to emit 
stochastically. And, as such, PAH molecules within the enshrouding dust shell will not emit significantly 
since any radiation from the nuclear power source(s) will be devoid of UV/optical photons by the time it reaches them. 
Of course, it is possible that additional ``knots'' of star-formation exist within such an obscuring shell---the models presented 
here all assume a central nuclear power source located within the shell, and thus do not address this possibility.

For the distribution of dust within the shell, we adopt a radial power law, $\rho \propto r^{-p}$, with $p$ a free parameter. As shown
in Figure~\ref{fig:ULIRGModel}, the shell extends from an inner radius $r_{\rm in}$, where the dust temperature is \Tin, to an outer 
radius $r_{\rm out} = Y r_{\rm in}$ (i.e., $Y$ is the relative thickness). The actual values of $r_{\rm in}$ and $r_{\rm out}$ never 
enter---only \Tin\ and $Y$ are relevant, and are specified as input parameters. Specifying the V-band dust optical depth through 
the shell, \tV, sets the optical depths at all other wavelengths. Given the SED of the input heating source, 
the solution to our radiative transfer problem is therefore fully determined by the values of only four parameters: 
(1) the relative thickness of the obscuring dust cloud, $Y$; 
(2) the exponent, $p$, of the radial dust density distribution; 
(3) the total $5500\Angstrom$ ({\it V}-band) optical depth, \tV, to the heating source; and
(4) the thermal equilibrium temperature, \Tin, of the dust at the inner-edge of the obscuring cloud.

We have generated models of ULIRGs with obscuring dust shells of relative thicknesses $Y = 100$, 200, and 400 
and radial density distributions of $p = 0$, 1, and 2. These choices permit us to investigate a wide range of values 
spanning the reasonable parameter space for an obscuring nuclear shell. Our models range from optically thin 
shells ($\tV \ll 1$) up to \tV\ = 200. Assuming standard interstellar dust abundance, the shell mass is 
$\E7\Mo\x\tau_{\rm V,100} L_{12}Y_{100}k_p$, where $\tau_{\rm V,100} = \tV/100$ 
and $Y_{100} = Y/100$. The value of $k_p$ is $30Y_{100}$, $11Y_{100}/(1 + 0.2\ln Y_{100})$, and 1 for 
$p$ = 0, 1, and 2, respectively. 

With these choices made, we have only to decide upon the assumed equilibrium temperature of dust at the inner-edge 
of the obscuring shell. As depicted in Figure~\ref{fig:ULIRGModel}, the inner edge of the shell in our purely 
AGN-powered model is assumed to be coincident with the clumpy torus surrounding the AGN accretion disk.
We assume that the hard radiation from the AGN accretion disk therefore dominates the heating of the obscuring shell 
and raises its inner edge up to the grain sublimation temperature. Thus, for the case of pure AGN heating, we assume
that the dust temperature at the inner-edge of the shell,  \Tin, is equal to our adopted grain sublimation temperature
of 1,000\,K. Note that we found little variation in our results for slightly higher assumed sublimation temperatures. 
For reference, this choice of grain sublimation temperature gives a distance of $r_{\rm in} = 2.3\,L_{12}^{1/2}$ pc 
to the inner edge of the obscuring shell in our model (see eq.~[1] from \citealt{Nenkova08}). 

For ULIRGs not powered entirely by an AGN, the choice of the dust shell's inner edge temperature is 
less straightforward since the heating source changes from central to fully distributed for a pure starburst 
(with a combination of both for an AGN/starburst hybrid).
Motivated by our desire to build and test a simple yet comprehensive ULIRG model, we assume that 
although the nature of the heating source changes in going from a pure AGN to a pure starburst-powered system, 
the geometrical properties of the obscuring shell---in particular, $r_{\rm in}$, its physical size---remains constant. 
This choice can be justified if we assume that: (1) the vast majority of ULIRGs are powered by both star-formation 
and AGN activity at various points in their histories; (2) the relative strengths of these power sources vary over time; and 
(3) every ULIRG is AGN-dominated at some point in its history (perhaps just in its nascent phase before significant 
star-formation begins) even if no AGN activity is currently detectable.

When the ULIRG is in its AGN-dominated phase, dust in close proximity to the accretion disk will sublimate 
and the inner edge of the obscuring shell will therefore be coincident with the inner edge of the AGN torus. 
As noted above, for a galaxy of a given luminosity, this assumption sets the physical size of the obscuring shell's inner 
edge ($\sim$2.3 pc for an $\Lbol = 10^{12}\LSun$ AGN). Dusty knots of star-formation cannot exist within the cavity 
bounded by the shell's inner edge since dust in this region would also sublimate. Thus, as depicted in 
Figure~\ref{fig:ULIRGModel}, we assume that star-formation in a ULIRG's nucleus is distributed around the 
inner edge of the obscuring shell. In other words, AGN activity clears the cavity and sets its 
physical dimensions---any subsequent changes in star-formation and/or AGN activity within or along the inner 
edge of the shell has no effect on the shell's physical size, but it does change the spectrum of radiation which heats it. 
While these assumptions certainly do not hold for all ULIRGs, it is reasonable to assume that they hold for a 
sufficiently high fraction of sources to justify their adoption in our modeling.

In this scenario, a pure starburst's infrared-dominated SED heats a surrounding shell less efficiently than the 
UV/optical-dominated SED of a pure AGN (since both are at the same physical distance from the dust\footnote{Note 
that dust along the inner edge of the shell receives the same flux (so is heated to the same temperature) whether stars 
are symmetrically distributed around the edge of the cavity or all placed at its center. This means that the inner edge 
dust temperature can be calculated by assuming that all flux from the combined heating spectrum is emitted at the 
center of the cavity.}). For pure starburst heating, the relatively unobscured starburst in our model 
raises the shell's inner edge to a temperature of \Tin\ = 580\,K, while our obscured starburst heating source 
(with less UV/optical emission) raises the shell's inner edge temperature to just \Tin\ = 350\,K. Mixtures of the input heating 
templates shown in Figure~\ref{fig:InputSEDs} produce intermediate dust temperatures: AGN fractions of 75\% and 50\%
yield \Tin\ = 980\,K and 885\,K for both the low- and high-obscuration starbursts, while an AGN fraction of 25\% yields 
\Tin\ = 760\,K and 735\,K for the low- and high-obscuration starbursts, respectively.

\subsection{ULIRG Nucleus Emission}
\label{sec:ModelingNuclearEmission}

\begin{figure*}
  \epsscale{0.75}
  \plotone{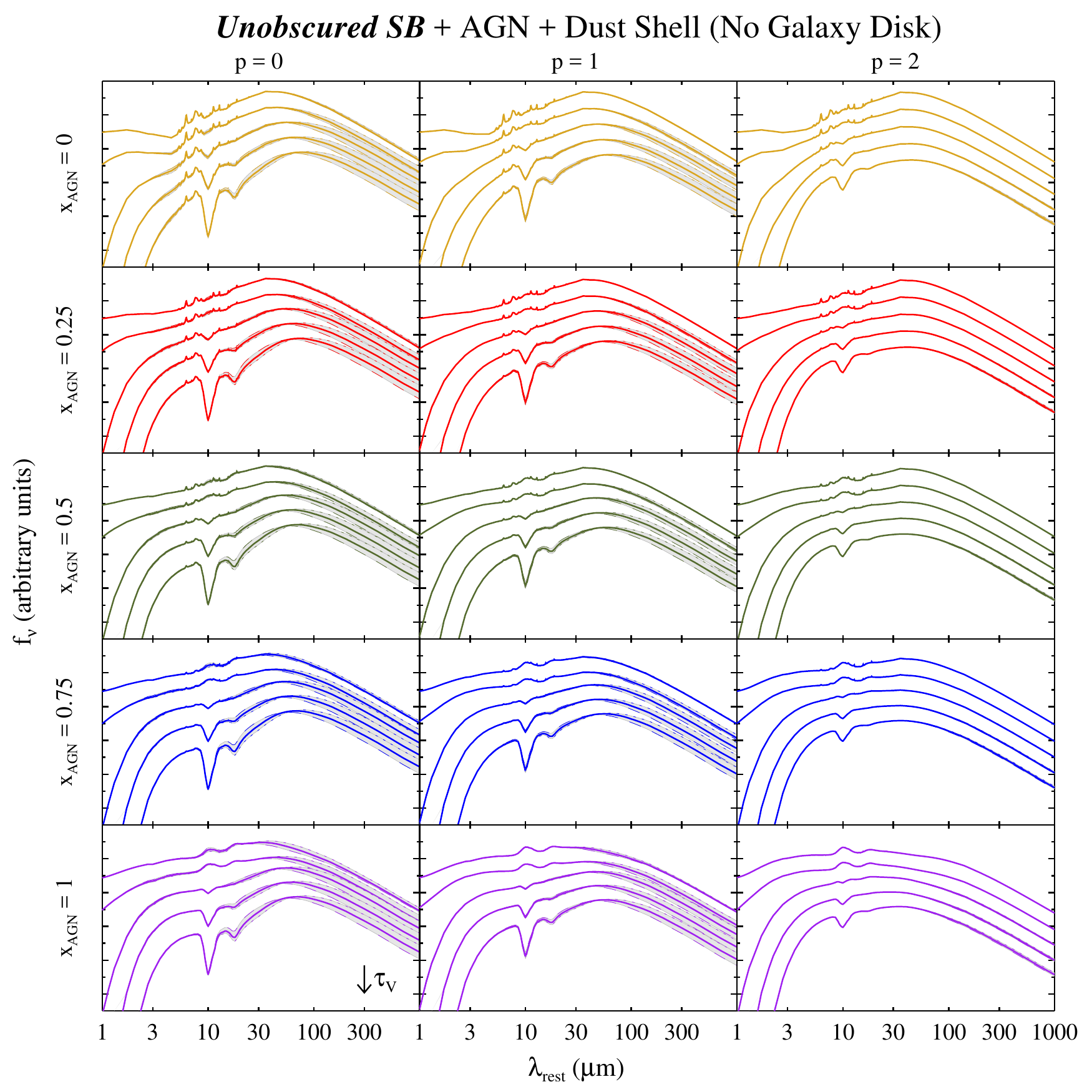}
  \plotone{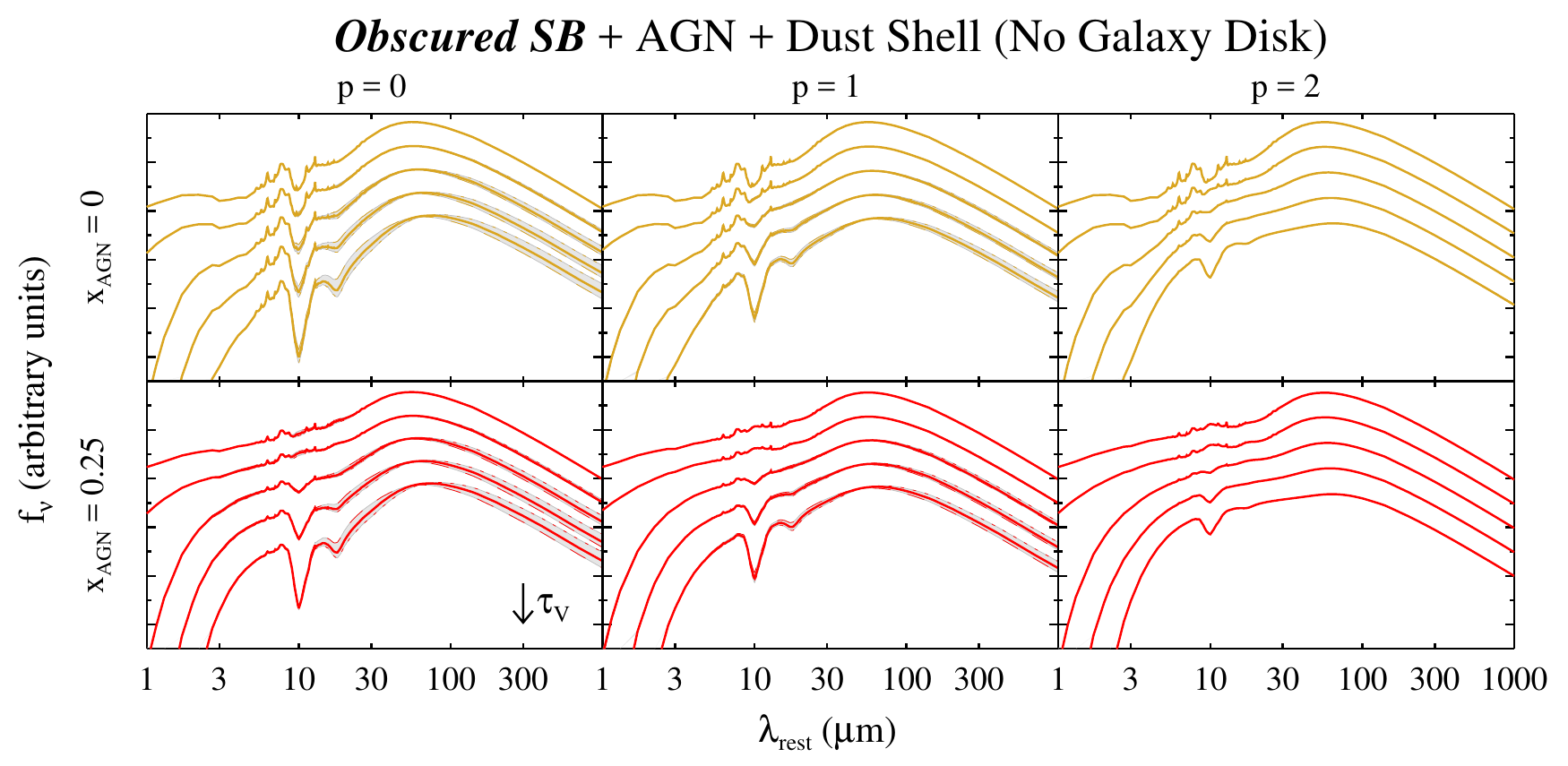}
 
\caption{Infrared SEDs of our obscured ULIRG nuclei models containing emission from the 
attenuated central engine and the obscuring dust shell (but not a host galaxy disk as in Fig.~\ref{fig:ModelSEDs_IR}). 
SEDs are shown for central engine heating sources comprised of an AGN and either a relatively unobscured
starburst nucleus (\textit{top}) or obscured starburst nucleus (\textit{bottom}). Each row corresponds to heating 
by the AGN/starburst mixture with the indicated \xAGN\ value, and each column corresponds to the indicated value 
of the dust shell radial density distribution index,  $p$. From top-to-bottom in each sub-panel, curves represent models 
with shell optical depths of $\tV = 1$, 10, 50, 100 and 200. All models are normalized to have unit integrated flux density, 
and the SEDs of models with increasing \tV\ are each scaled to be a factor of 10 smaller than the previous (for clarity). 
All models are shown for obscuring dust shell thicknesses of $Y = 100$, 200, and 400. The $Y = 200$ model is shown 
with a solid thick line and the flanking thin lines (and shaded region between them) show the model's range between 
$Y = 100$ (\textit{dashed lines}) and $Y = 400$ (\textit{dotted lines}). These models assume a fully covered central 
source (i.e., $\xShell = 1$ in eq.~[\ref{eq:f_nucleus}]). Models for the obscured starburst heating source are only shown 
for $\xAGN = 0$ and 0.25 since they are indistinguishable from the unobscured heating source case for all other 
\xAGN\ values (and even nearly indistinguishable---except for small variations for models with different $Y$ 
values---for the $\xAGN = 0.25$ case). The most noticeable differences between the unobscured and obscured 
starburst heating source models are seen in the low-\tV, $\xAGN = 0$ cases. For $\tV \ga 50$ in the $p = 1$ and 2 
cases, and $\tV \ga 100$ in the $p = 0$ case, noticeable differences between the SEDs produced by obscured and 
unobscured starburst heating sources disappear.}

  \label{fig:ModelSEDs_NoDisk}
  \epsscale{1.0}
\end{figure*}

Our obscured ULIRG nucleus models include emission from both an obscured central heating source having 
intrinsic (i.e., unobscured) spectral shape, $f_\nu^\mathrm{source}$, and reprocessed radiation from a
surrounding obscuring dust shell, $f_\nu^{\rm shell}$. In constructing our models of the emission from the ULIRG nucleus,
we allow for the possibility that some radiation from the obscured central engine can escape unimpeded through 
``keyhole'' passages in the shell, so that only a fraction, $0 \le \xShell \le 1$, of the central radiation is captured and 
reprocessed. The remaining fraction, $1 - \xShell$, emerges unchanged. In the most general case then, the spectral 
shape of the radiation emerging from the nucleus is itself a mixture:
\begin{equation}
   \label{eq:f_nucleus}
   f_\nu^\mathrm{nucleus} = (1 - \xShell) f_\nu^\mathrm{source} +
                                              \xShell f_\nu^{\rm shell}
\end{equation}
In this equation, all flux densities are normalized such that $\int f_\nu \,d\nu = 1$. Note that both $f_\nu^\mathrm{source}$ 
and $f_\nu^\mathrm{shell}$ depend upon the AGN-fraction of the central heating source, $0 \le \xAGN \le 1$. 

The combined radiation emerging from the nucleus, $f_\nu^{\rm nucleus}$, is shown in Figure~\ref{fig:ModelSEDs_NoDisk} 
for different combinations of central engine heating sources and dust shell properties. Note that models 
in Figure~\ref{fig:ModelSEDs_NoDisk} are for fully covered geometries ($\xShell = 1$) and do not include 
emission from a host galaxy disk. Related models with host galaxy disk emission are shown in 
Figure~\ref{fig:ModelSEDs_IR}, and their relationship to the models in Figure~\ref{fig:ModelSEDs_NoDisk} are 
discussed in \S\ref{sec:ModelingTotalEmission}.

\subsection{Galaxy Disk Emission}
\label{sec:ModelingDiskEmission}

\begin{figure}
  \epsscale{1.23}
  \plotone{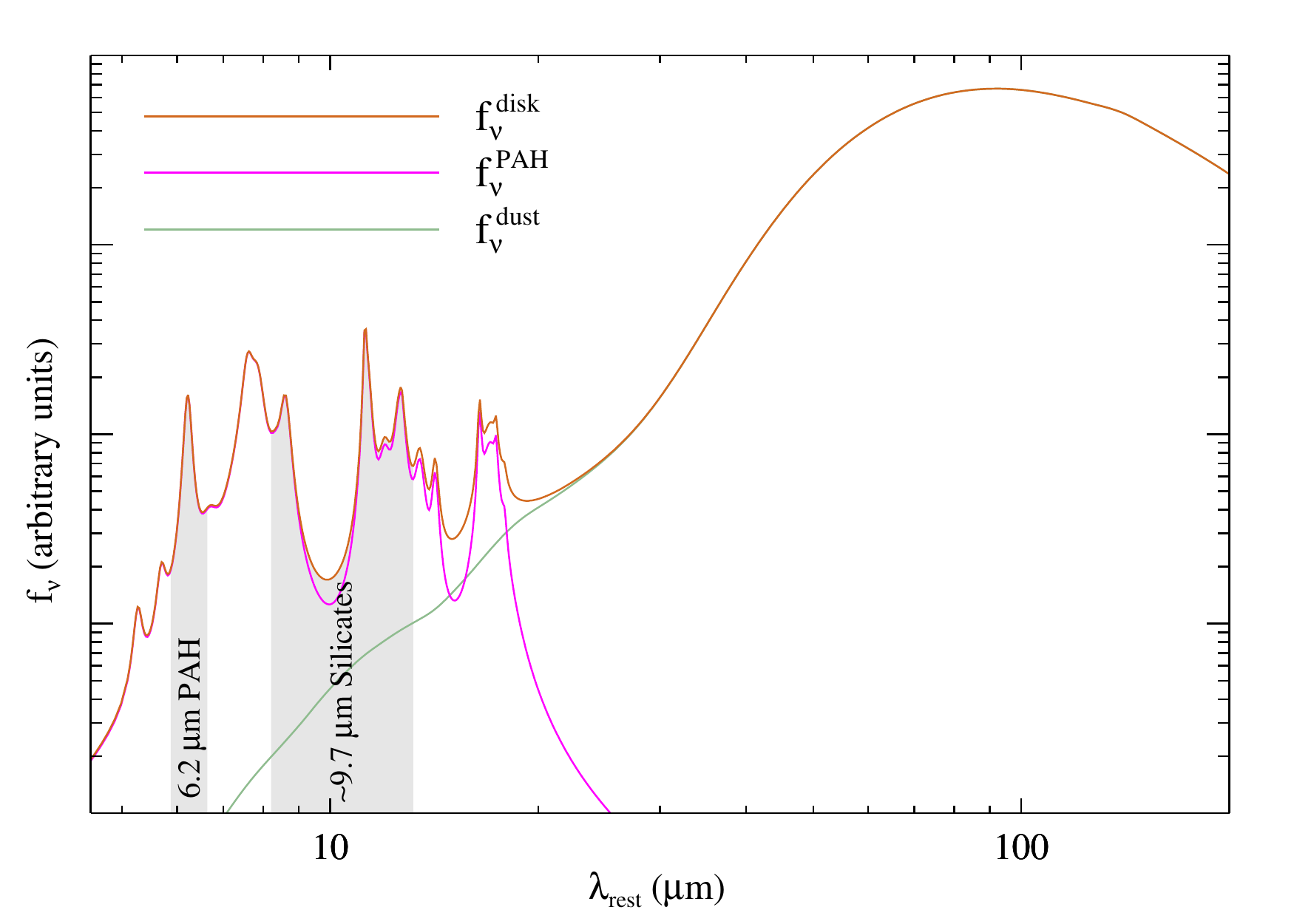}
 
\caption{The SED of the host galaxy disk component used in our modeling is based upon the
 ``global'' fit to the observed SED of our unobscured starburst prototype, NGC\,7714, presented in  
 \citet{Marshall07}. The complete disk model is shown along with emission from the mid-IR PAH and 
mid-to-far-IR dust sub-components from which the model is built.}
  \label{fig:InputSEDs_Disk}
  \epsscale{1.0}
\end{figure}

Our model also includes a disk emission component which represents unobscured dust and 
PAH emission from the extended disk of a ULIRG's host galaxy. This model, shown in 
Figure~\ref{fig:InputSEDs_Disk}, is composed of far-IR thermal dust continuum emission
and mid-IR PAH feature emission:
\begin{equation}
   \label{eq:f_disk}
      f_\nu^\mathrm{disk} = \xPAH f_\nu^{\rm PAH} + (1 - \xPAH) f_\nu^\mathrm{dust}
\end{equation}
In this equation, all flux densities are normalized such that $\int f_\nu \,d\nu = 1$ and 
$\xPAH \equiv L_\mathrm{PAH} / (L_\mathrm{dust} + L_\mathrm{PAH})$ is the fraction
of the disk component luminosity attributable to PAH emission. For our models, we adopt 
$\xPAH = 0.1$, a value typical for local star-forming galaxies \citep{Smith07}.

The normalized flux densities of the model's two sub-components, $f_\nu^{\rm PAH}$ and
$f_\nu^\mathrm{dust}$, are derived from the ``global'' fit to NGC\,7714 from \citet{Marshall07}. This fit was 
performed using large aperture far-IR photometric data and therefore contains significant mid-to-far-IR PAH 
and dust emission from the host galaxy disk. We adopt the PAH spectrum of NGC\,7714 from that fit for 
our model of the emission from typical unobscured PAHs in host galaxy disks, $f_\nu^\mathrm{PAH}$.
We use the ``cold'' ($T \approx 30\K$), ``cool'' ($T \approx 70\K$), and ``warm'' ($T \approx 165\K$) dust 
components from the fit in \citet{Marshall07} to construct our model of the mid-to-far-IR thermal dust 
continuum, $f_\nu^\mathrm{dust}$. The fitted ``cool'' and ``warm'' components make up 5\% of the total 
dust luminosity in our model, and are included to provide a low-level continuum beneath the PAH features 
and to match the observed mid-IR SEDs of regions within typical galactic disks---e.g., the 
$f_\nu(20\um) / f_\nu(5.5\um) \sim 4$ of our model is similar to the slope observed in the coolest 
(in terms of mid-IR slope) regions of the M31 disk \citep{2015MNRAS.454..818H}.

\subsection{Total ULIRG Emission}
\label{sec:ModelingTotalEmission}

\begin{figure*}
  \epsscale{0.75}
  \plotone{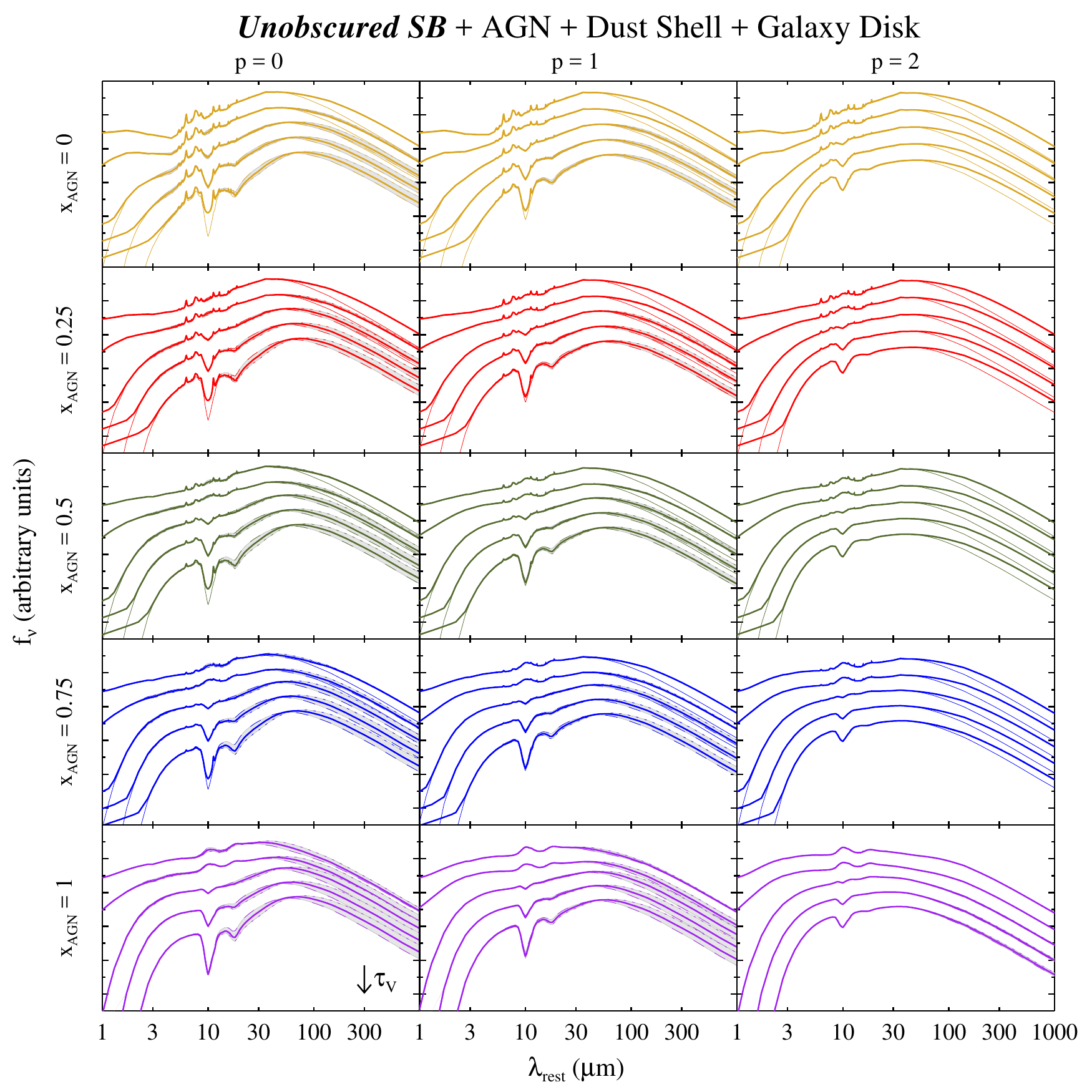}
  \plotone{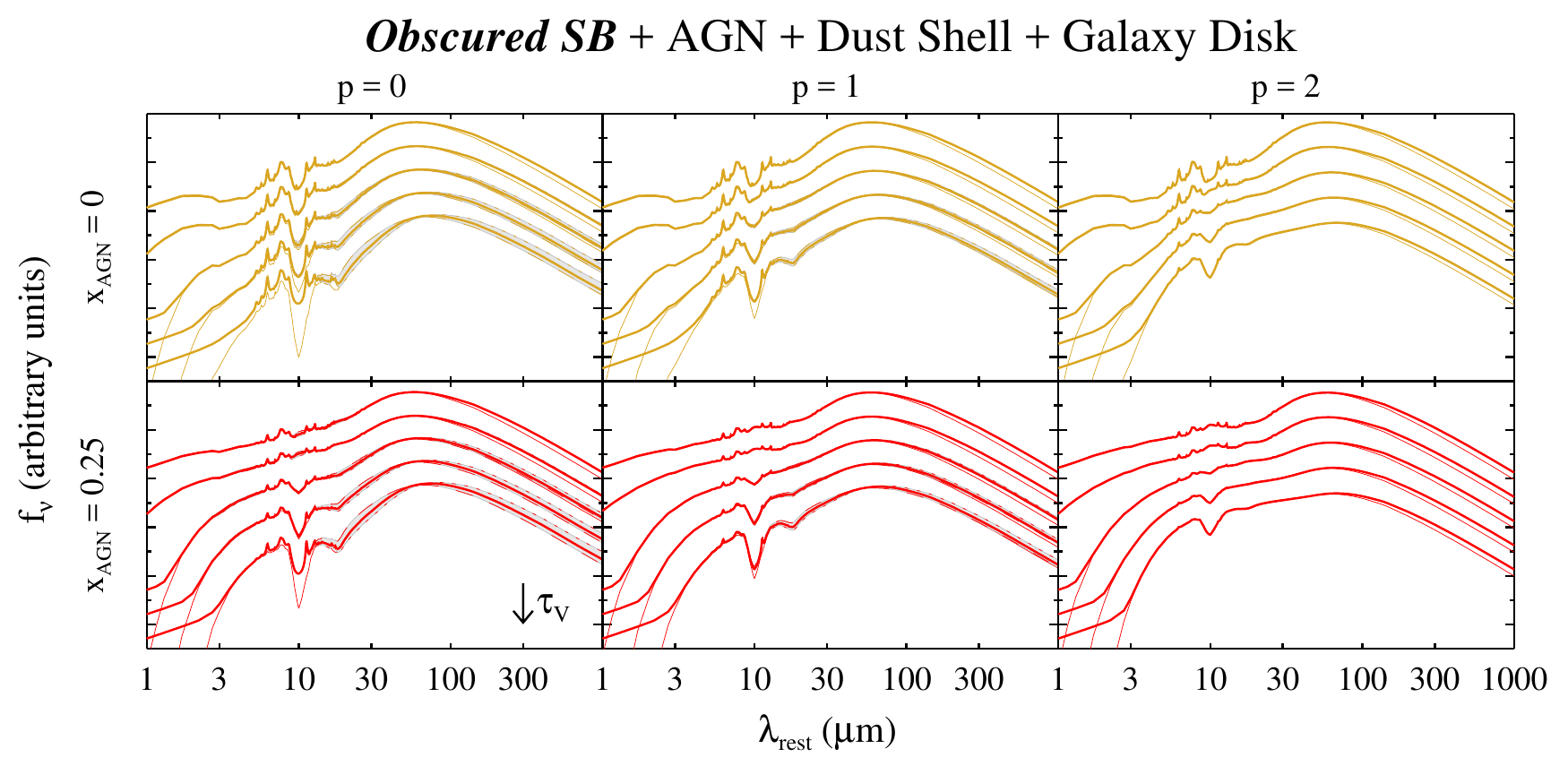}
   
\caption{Near-to-far-IR SEDs of the total emission from our models of an obscured ULIRG nucleus and
its host galaxy's disk. The obscured nuclear component includes emission from the attenuated power source and
the surrounding obscuring shell. The power source itself is composed of emission from our prototype AGN and
either our unobscured (\textit{top}) or obscured (\textit{bottom}) prototype starburst. These models are as described 
in Fig.~\ref{fig:ModelSEDs_NoDisk}; but, unlike those ($\alpha_\mathrm{disk} = 0$) models with no disk emission, 
these $\alpha_\mathrm{disk} = 0.1$ (see eq.~[\ref{eq:L_disk}]) models explicitly contain emission from the host galaxy 
disk. Once again, models for the obscured starburst heating source are only shown for $\xAGN = 0$ and 0.25 since 
they are indistinguishable from the unobscured heating source case for all other \xAGN\ values. Of particular interest 
is the fact that the small amount of unobscured PAH emission from the host galaxy disk in these models quickly 
``fills-in'' the deep $\sim9.7\um$ silicate features seen in Fig.~\ref{fig:ModelSEDs_NoDisk}. To make this easier to 
see, the $Y = 200$ ``no-disk'' models from Fig.~\ref{fig:ModelSEDs_NoDisk} are shown as thin solid lines. The 
host galaxy disk contribution is most noticeable in the near-IR and beneath the $\sim9.7\um$ silicate feature 
(due to PAH emission), as well as in the far-IR (due to thermal dust emission).}

  \label{fig:ModelSEDs_IR}
  \epsscale{1.0}
\end{figure*} 

\begin{figure*}
  \epsscale{0.75}
  \plotone{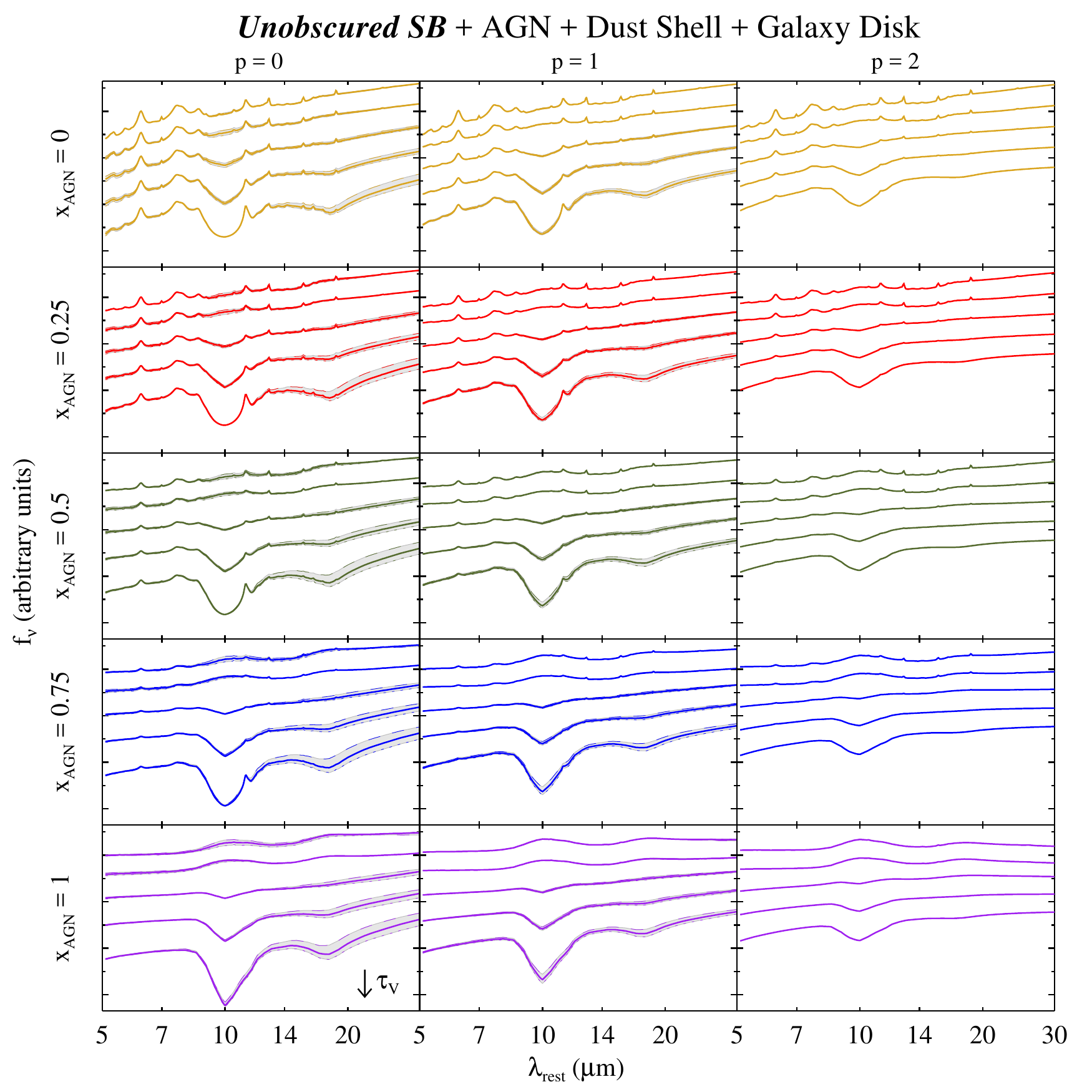}
  \plotone{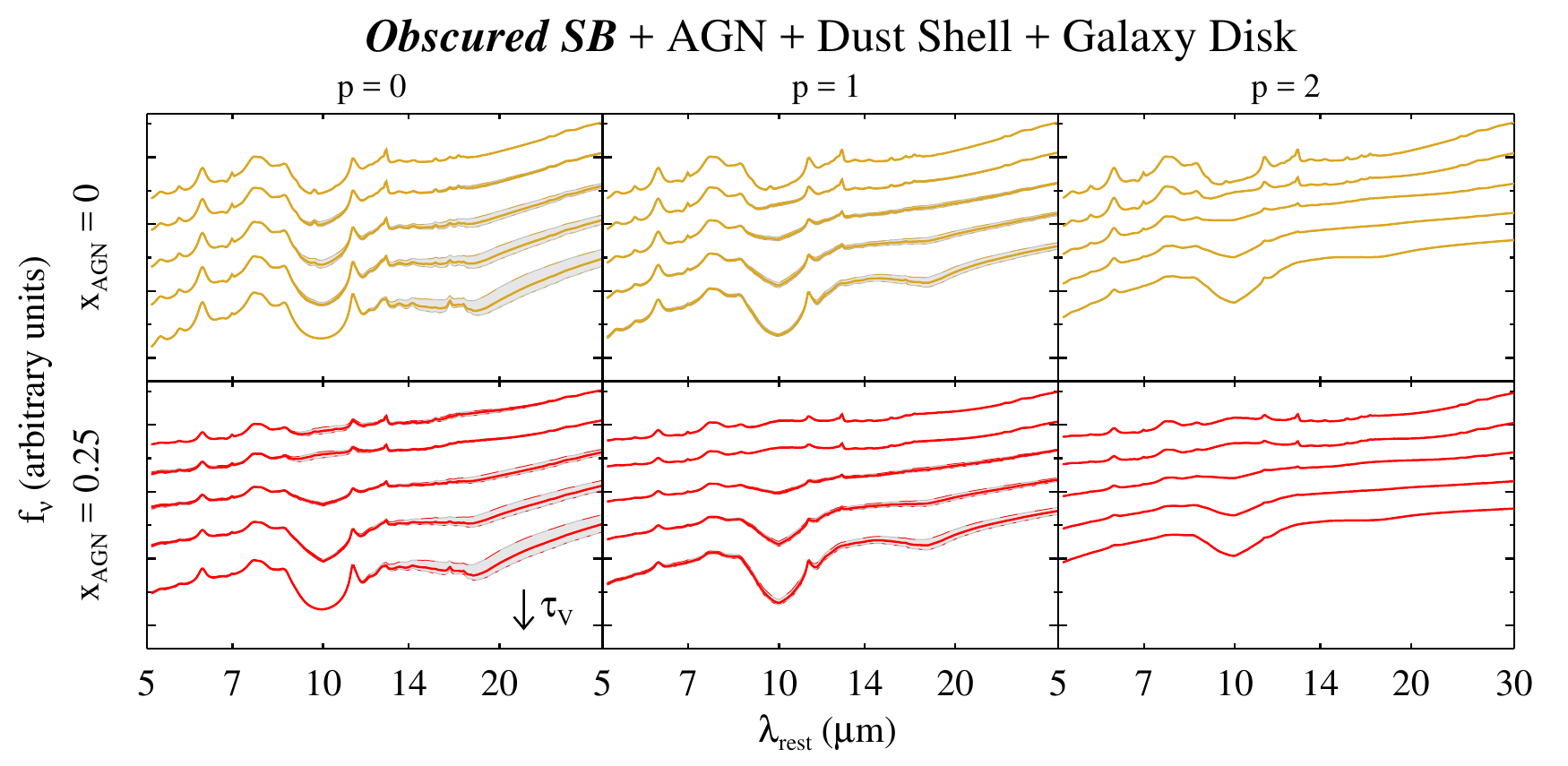}
   
\caption{Close-up mid-IR views of the SEDs shown in Fig.~\ref{fig:ModelSEDs_IR} for our model ULIRGs
powered by emission from our prototype AGN and either our unobscured (\textit{top}) or obscured (\textit{bottom}) 
prototype starburst. Note the remarkable resilience of the PAH features. In the models with pure starburst heating 
($\xAGN = 0$), the features are clearly discernible even after reprocessing by $\tV = 200$ shells. See \S\ref{sec:PAH} 
for a discussion of this phenomenon.}

  \label{fig:ModelSEDs_MIR}
  \epsscale{1.0}
\end{figure*} 

\begin{figure*}
  \epsscale{1.17}
  \plottwo{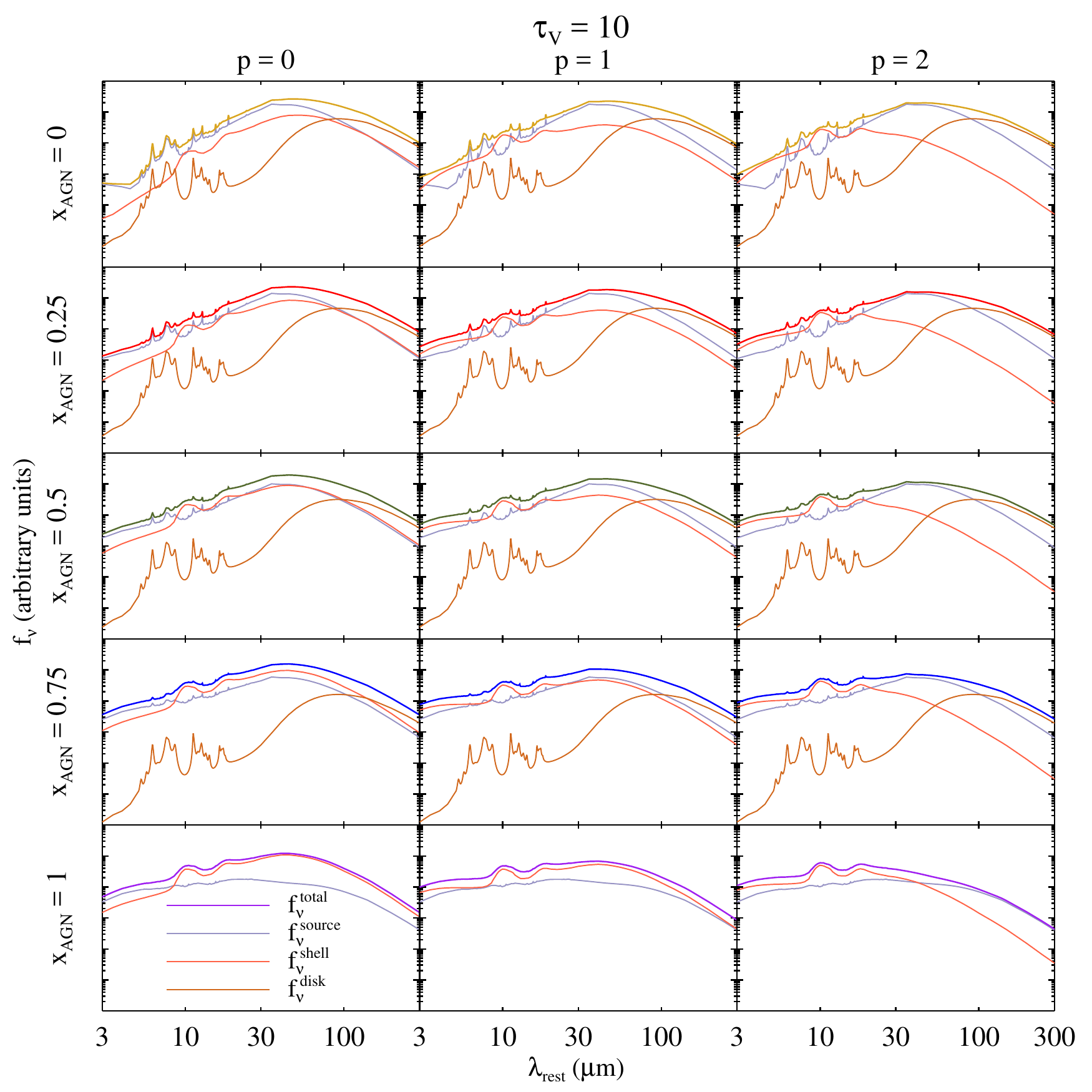}{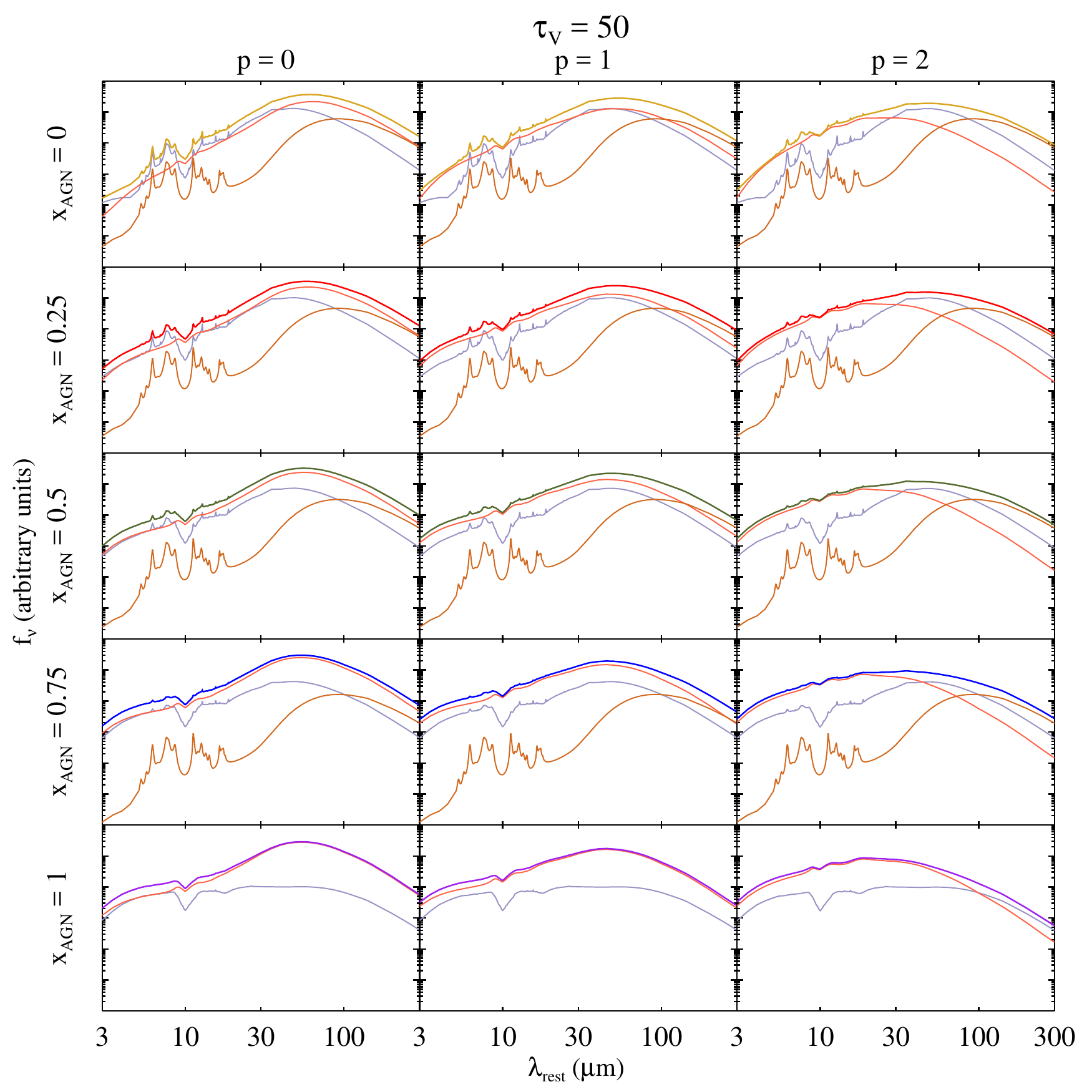}
  \plottwo{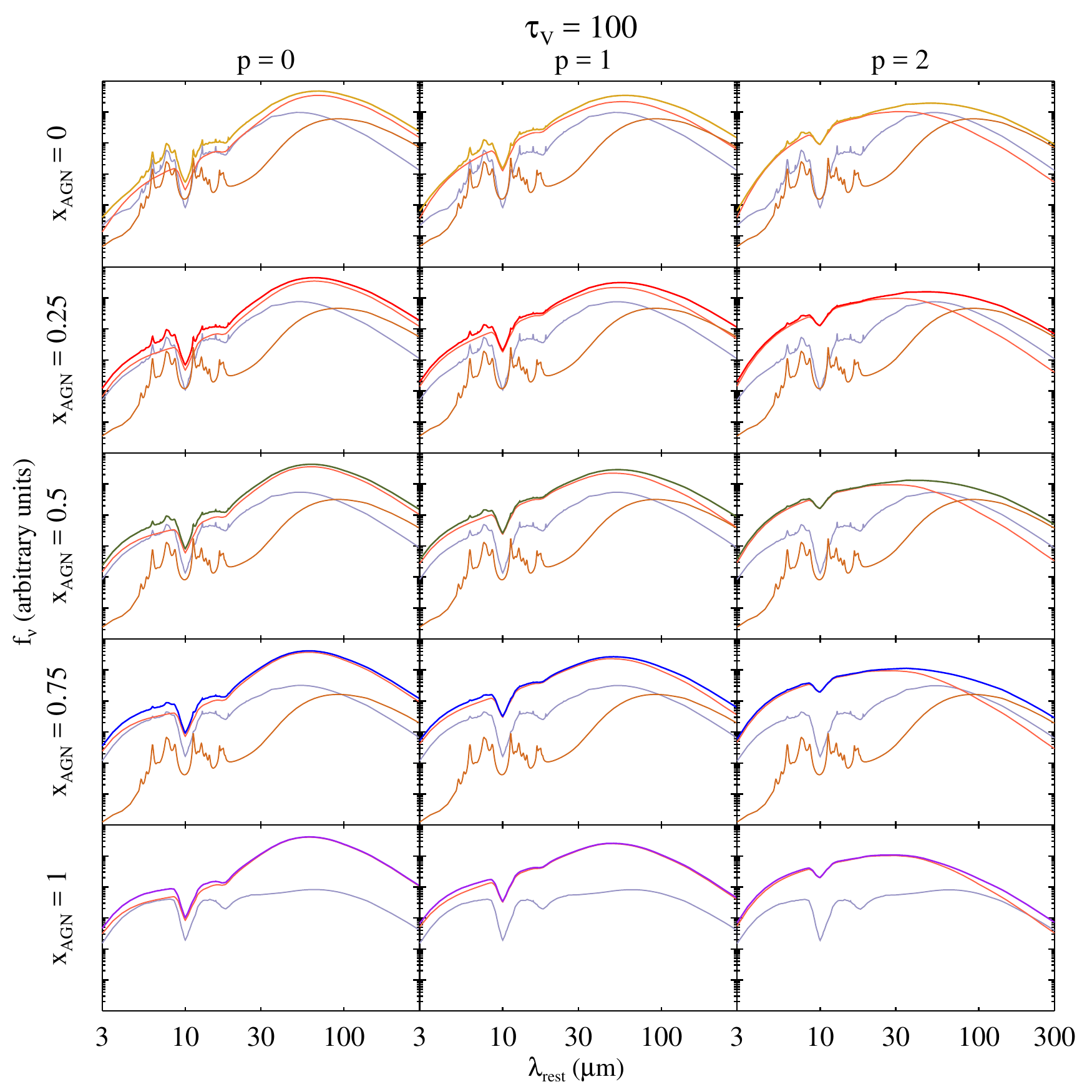}{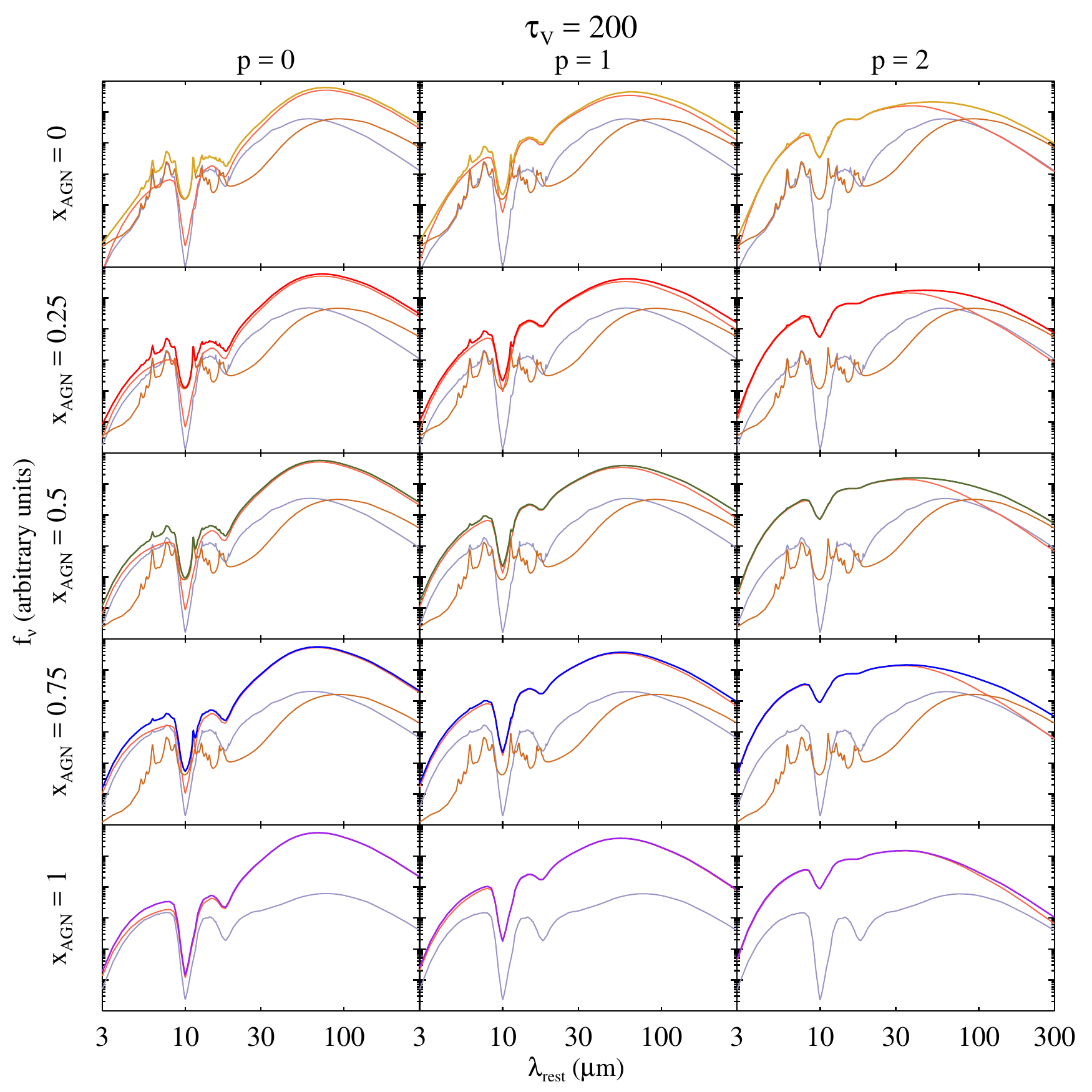}
 
\caption{The ``source'' (\S\ref{sec:ModelingCentralEngine}), ``shell'' (\S\S\ref{sec:ModelingShellEmission} 
and \ref{sec:ModelingNuclearEmission}), and ``disk'' (\S\ref{sec:ModelingDiskEmission}) components of the 
unobscured starburst and AGN-powered ULIRGs from Fig.~\ref{fig:ModelSEDs_IR} are shown for various values 
of \xAGN\ at $\tV = 10$, 50, 100, and 200 as labeled. $Y = 200$ for all models and $p = 0$, 1, and 2 as indicated 
for each column.}

  \label{fig:ModelSEDs_Components}
  \epsscale{1.0}
\end{figure*}

The models described in \S\ref{sec:ModelingNuclearEmission} represent the emission from obscured ULIRG 
nuclei. But the observed SED of an entire ULIRG is composed of both this obscured nuclear emission and the 
relatively unobscured emission from its host galaxy disk. At the distance to the median
galaxy in our observed sample (to be described in \S\ref{sec:AnalysisModelsData}), the SL slit of the
\Spitzer\ \IRS\ instrument spans a physical projected size of $\sim$10.6~kpc. The portion of the \Spitzer\ \IRS\ spectra 
of galactic nuclei used in our analysis must, therefore, include a significant contribution from the host galaxy disk. 
We therefore model the total emission from a ULIRG as the sum of its nuclear emission, $F_\nu^\mathrm{nucleus}$,
and the emission from its host galaxy disk, $F_\nu^\mathrm{disk}$.

The monochromatic luminosity of each component can be written as $L_\nu = L f_\nu$, where $L$ is the
component's bolometric luminosity and $f_\nu$ its normalized spectral shape, 
$f_\nu \equiv F_\nu / \int\!\! F_\nu d\nu$. Therefore, the spectral shape of the overall
emergent radiation comprised of a mixture of the nuclear and disk components can be written
\begin{equation}
   \label{eq:TotalEmissionModel}
   f_\nu^\mathrm{total} = \frac{L_\mathrm{nucleus} f_\nu^\mathrm{nucleus} +
                                L_\mathrm{disk} f_\nu^\mathrm{disk}}
                               {L_\mathrm{nucleus} + L_\mathrm{disk}}
\end{equation}

The luminosities of the two components,  $L_\mathrm{nucleus}$ and $L_\mathrm{disk}$ only 
enter our models through the ratio
\begin{equation}
   \label{eq:L_disk}
   \alpha_\mathrm{disk} \equiv \frac{L_\mathrm{disk}}{(1 - \xAGN) L_\mathrm{nucleus}},
\end{equation}
where $(1 - \xAGN) L_\mathrm{nucleus}$ is the starburst component of
the nuclear luminosity. $\alpha_\mathrm{disk}$ is therefore a parameter that characterizes the 
luminosity of disk star-formation relative to that of the nucleus. \citet{DiazSan2010} show that 
$< 30\%$ of the $5$--$15\um$ flux of ULIRGs originates outside an unresolved point source, 
and that ULIRGs which are stage 4 advanced mergers have $< 20\%$ extended emission. Based 
upon these findings and additional results from our modeling described in \S\ref{sec:discussion}, 
we adopt a value of $\alpha_\mathrm{disk} = 0.1$ throughout this work (i.e., the luminosity of disk 
star-formation is 10\% the nuclear starburst luminosity). We therefore explicitly include contributions 
from unobscured mid-to-far-IR thermal and PAH emission in all models that have some starburst
contribution to the central engine SED (i.e., all models except $\xAGN = 1$).

Figures~\ref{fig:ModelSEDs_IR} and \ref{fig:ModelSEDs_MIR} show near-to-far-IR and mid-IR views of 
the total emission from our deeply buried ULIRG models, respectively. These models differ from those in 
Figure~\ref{fig:ModelSEDs_NoDisk} only in that they contain host galaxy disk emission. 
Figure~\ref{fig:ModelSEDs_Components} provides a detailed look at the ``source'' ($f_\nu^\mathrm{source}$),
``shell'' ($f_\nu^\mathrm{shell}$), and ``disk'' ($f_\nu^\mathrm{disk}$) components that contribute to the total 
SEDs ($f_\nu^\mathrm{total}$) of a subset of the models shown in Figures~\ref{fig:ModelSEDs_IR} 
and \ref{fig:ModelSEDs_MIR}. As will be discussed in \S\ref{sec:sil}, at large optical depths every model 
develops a 9.7\mic\ silicate absorption feature which gets deeper as \tV\ increases. For a given \tV, the feature 
depth decreases with $p$ and is relatively independent of $Y$ for the values explored in this work. Steep density 
distributions (such as $p = 2$) resemble thin shells with little temperature variation (since most of the dust is 
concentrated in a small radial range), and therefore have absorption features that tend to be shallow 
since the absorption feature is ``filled-in'' by hot dust emission from the thin shell. In contrast, 
radiation from the hot inner regions of obscuring shells with flat density distributions propagates through 
large quantities of cool dust, resulting in deep absorption features.

The exception to this is for models with starburst-dominated central engines in which PAH emission from 
the host galaxy disk ``fills-in'' the absorption feature and keeps it from increasing in depth (this PAH emission also 
has a pronounced effect on the near-IR emission of these models). In comparison, the models in 
Figure~\ref{fig:ModelSEDs_NoDisk} have no disk emission and thus have much deeper absorption features. 
The effect of this PAH emission on the spectra of our models is also apparent in the relative strengths of the
components shown in Figure~\ref{fig:ModelSEDs_Components}. This phenomenon wherein ULIRG SEDs are 
significantly affected by emission from their host galaxy disks is further discussed in \S\ref{sec:discussion}.

\section{Mid-IR Spectral Indicators}
 \label{sec:SpectralIndicators}

While a direct comparison of our models to a set of observed ULIRG SEDs could, in principle, be made by 
performing detailed fits to the data, such efforts have proven to be somewhat futile since the parameter space 
is not sufficiently well-constrained. We have instead found that a comparison of well-chosen spectral 
indicators from our models to the data is a more fruitful path forward. 

In particular, the 9.7\mic\ silicate feature strength ($S_{\rm sil}$), 
6.2\mic\ PAH feature equivalent width (${\rm EW}$), and $F_\nu(14\um) / F_\nu(5.5\um)$ 
continuum slope are the most prominent features in the mid-IR spectra of ULIRGs and thus serve as 
excellent proxies to characterize their entire mid-IR SEDs \citep[see][]{Laurent00, Spoon07, Armus07, Petric11}. 

We begin with a description of each of these features and a look at the phenomenology of how they arise 
in our suite of models and the spectra of dusty galaxies.

\subsection{$9.7\um$ Silicate Feature}
\label{sec:sil}

The most prominent feature in the mid-IR spectra of many dusty galaxies occurs
as a result of absorption and/or emission by silicate grains at approximately
$9.7\um$. In quasars and Seyfert galaxies, this feature appears as either an
emission or a shallow absorption feature, whereas in ULIRGs that are not
identified as AGNs the feature often appears as a deep absorption feature \citep{Hao07}.

Assuming that $F_\nu^{\rm obs}$ and $F_\nu^{\rm cont}$ are the flux densities of the
observed spectrum and underlying continuum, respectively, the $9.7\um$ silicate 
feature strength of a spectrum is defined as
\begin{equation}
   \label{eq:Sil}
   \Sil = \ln{\frac{F_\nu^{\rm obs}(9.7\um)}{F_\nu^{\rm cont}(9.7\um)}}
\end{equation}
Positive values of $\Sil$ correspond to silicate emission features, while negative values
indicate absorption features. ULIRG absorption features range down to $\Sil \approx -4$ 
where the feature dips by more than a factor of 50 below the continuum. 

In principle, an absorbing cold dust screen between the observer and the IR emitter 
could produce such a deep feature. However, in order to remain cold, such a screen
could not also reprocess the amount of radiation indicated by the enormous \LIR\ 
values associated with ULIRGs. Thus, in addition to such a cold screen, we must 
also invoke an optically thick reprocessing component. Furthermore, the screen must 
always fully cover the primary reprocessing dust along the line of sight. Such a geometry 
requiring two spatially separate but correlated dust regions located along the line of sight 
represents a contrived solution for ULIRGs. In a more plausible and general explanation, 
a single entity of embedding dust must account for both the total IR luminosities and deep 
absorption features seen in ULIRGs.

The appearance of the silicate feature in the spectrum of radiation
reprocessed by dust was studied in detail by \citet{Levenson07} and \cite{Sirocky08}.
Radiation from centrally heated dust that is optically thin around 10\mic\
shows the feature in emission. As the optical depth through the obscuring
medium increases, the feature transitions into an increasingly deep absorption
feature. In addition to having a large optical depth through the obscuring
cloud, an absorption feature requires the presence of a temperature gradient
across the intervening dust---the larger the gradient, the deeper the feature.
Geometrically thin dust shells have a limited temperature range and therefore a
limited absorption depth (since their emission will effectively ``fill-in'' any
absorption feature). A deep feature therefore requires a central radiation
source that is enshrouded behind optically and geometrically thick dust so that radiation 
from the hot inner regions is reprocessed by successively distant outer 
layers containing increasing amounts of cooler dust. In addition, the density distribution must be smooth. 
Or, if the obscuring dust is clumpy, its volume filling factor must be very large to 
produce deep absorption. \citep[See also][where it is suggested that deep silicate absorption in some 
Compton-thick AGN may arise from dust in the host galaxy disk.]{2012ApJ...755....5G}
In contrast, dust in the obscuring torus of an AGN is clumpy and has a small volume filling factor 
so that an otherwise deep absorption feature in an AGN is ``filled-in'' by emission from the hot sides 
of clouds visible through the clumpy torus.

These conditions are fairly restrictive. Any significant deviation reduces the
depth of the absorption feature. While it is difficult to produce a deep
silicate absorption feature, it is easy to make a shallow one. As we will
describe in \S\ref{sec:discussion}, a mere 5\% mid-IR contamination from
low-obscuration star-formation is sufficient to ``fill-in'' an otherwise deep absorption feature. 
The infrared radiation from ULIRGs with deep silicate absorption features must, therefore, 
arise from a single deeply obscured central source with very little emission contributed
from elsewhere. The radiation emerging from this deep 
obscuration must also contain a sufficiently strong warm component to explain the observed flat mid-IR
slopes of some ULIRGs---a significant challenge. 

\begin{figure*}
  \epsscale{0.7}
  \plotone{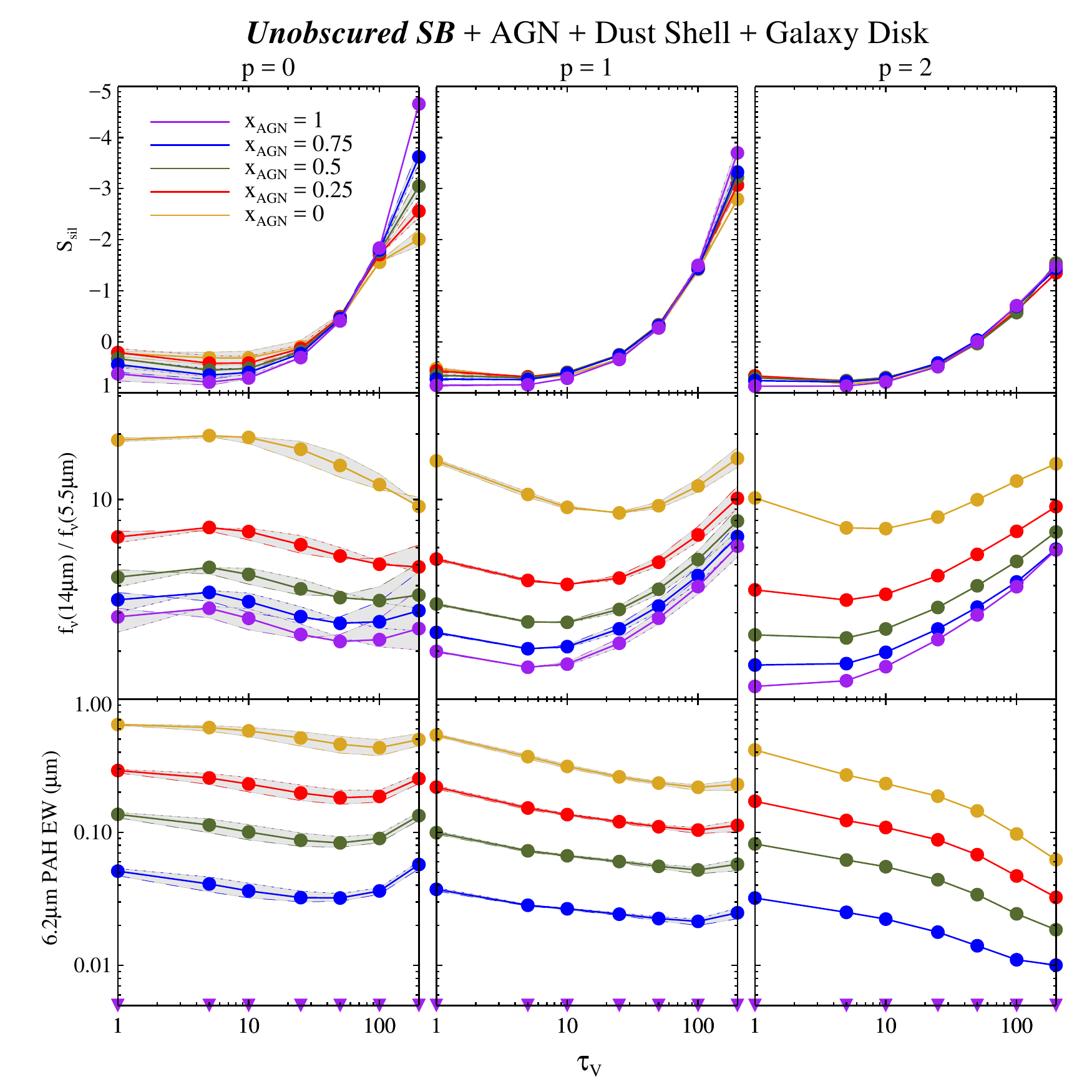}
  \plotone{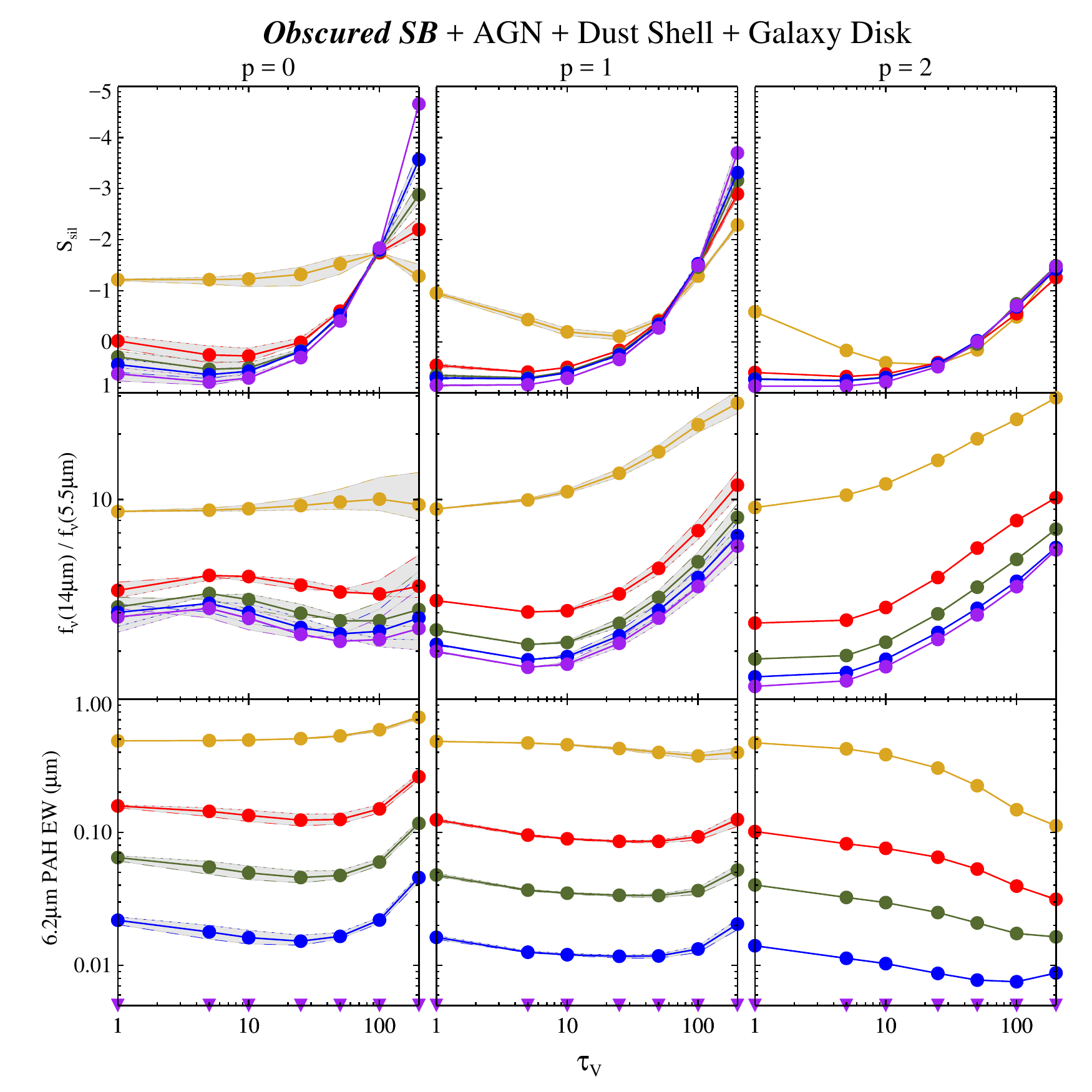}

\caption{Silicate feature strength ($S_{\rm sil}$), 14-to-5.5\mic\ continuum slope, and 6.2\mic\ PAH equivalent 
width as a function of optical depth through the obscuring shells ($\tV = 1$, 5, 10, 25, 50, 100, and 200 from 
left-to-right in each sub-panel) of the models shown in Fig.~\ref{fig:ModelSEDs_IR}. Spectral indicators are shown 
for all \xAGN\ and $p$ values as labeled. The $Y = 200$ model is indicated by a solid thick line and the flanking thin 
lines (and shaded region between them) show the model's range between $Y = 100$ (\textit{dashed lines}) and 
$Y = 400$ (\textit{dotted lines}). Downward arrows indicate PAH equivalent width upper-limits for the pure AGN models.}
  \label{fig:ModelParams}
  \epsscale{1.0}
\end{figure*}

\subsection{PAH Features}
\label{sec:PAH}

Beginning with the pioneering studies of Aitken, Roche and collaborators (see
\citealt{Roche91}, and references therein), it has been established that the
mid-IR spectra of starburst galaxies show prominent PAH emission features,
whereas AGN SEDs are generally devoid of such features. The latter has been
attributed to the destruction of the PAH molecules by AGN X-rays
\citep{Voit92}, although it may also be explained in terms of the feature
excitation mechanism. Namely, the features arise from the stochastic heating of
small grains that leads to sporadic deviations from equilibrium. Significant temperature 
spikes occur when the absorption of a single photon markedly increases a grain's internal 
energy, and PAH emission occurs during the subsequent cooling-down period 
\citep[see, e.g.,][]{Draine03}. This requires a combination of very small grain mass and 
a diffuse radiation field of high-energy (UV) photons. When any of these conditions is 
violated, the grains either sublimate or reach thermal equilibrium with the radiation field 
(whatever their size), in which case their emission shows only small fluctuations around 
black-body radiation at their equilibrium temperature.

In particular, temperature fluctuations of small grains heated by UV radiation
decrease when the photon density is sufficiently high that the grains are repeatedly hit
before they have time to cool significantly between photon absorptions. 
As a result, PAH emission from young, hot stars originates at
relatively large radial distances (where the photon density is sufficiently
low). For example, around hot O-stars with luminosities from
3\x\E3--5\x\E5\,\Lo, the PAH emission zone is at a radial distance of
\about0.2--1 pc, where the radiation field has been diluted by \about\E{-13}
\citep[e.g.,][]{Verstraete01}. Inside that radius dust emission is thermal.
Scaling to AGN luminosities of \E{12}\,\Lo\ yields radial distances of $10^2$--$10^3$\,pc, 
explaining why the central kiloparsec of an AGN host galaxy is a PAH
``zone of avoidance'' \citep{Desert88}. Strong PAH emission from starburst galaxies arises 
from many star forming regions distributed over a large volume, where the collective emission 
reflects the combined effect of many diffuse radiation regions.

Pertinent to this work, it follows that a collection of hot stars inside an optically thick dust shell heated to 
temperatures that generate mid-IR emission will not produce significant PAH features. When the dust shell's 
inner regions are heated to the high temperatures necessary for deep silicate absorption and appreciable mid-IR
emission, the photon density of the radiation field is sufficiently high that
all grains, including small surviving ones, are in thermal equilibrium. In the
dust shell's outer regions where the radiation field does become sufficiently diffuse, it is also
nearly devoid of UV photons (which are absorbed in the inner layers), and thus cannot
induce the stochastic heating necessary for PAH excitation. The presence of strong PAH features in 
ULIRGs therefore requires distributed---not centralized---heating. In contrast, as mentioned
in \S\ref{sec:sil}, deep silicate absorption features require central heating of dust that is both optically and
geometrically thick in order to provide a sufficient temperature gradient and cooler outer layers. 

To understand the relationship between the intrinsic and observed strength of
PAH emission as it propagates through an obscuring cloud (i.e., as expected in
a ULIRG), let's denote by $EW_0$ the equivalent width of a PAH feature
superposed on an underlying continuum with intensity $I_0$. When this radiation
propagates through dust with optical depth $\tau$ at the feature wavelength,
the attenuated intensity of the emerging radiation is $I_0 e^{-\tau}$, while
reprocessing of the radiation absorbed by the obscuring dust leads to thermal
emission with intensity $\Is$. Reprocessed radiation has no UV photons, and thus does not 
produce PAH emission. As a result, the emerging radiation exhibits a PAH feature with an equivalent width
\begin{equation}\label{eq:EW}
    EW(\tau) = EW_0\,{I_0 e^{-\tau} \over \Is + I_0 e^{-\tau}}
\end{equation}
As long as the reprocessed emission is small ($\Is \ll I_0 e^{-\tau}$), the
feature emerges with equivalent width $EW_0$. Once the reprocessed emission
becomes significant, the equivalent width is reduced, decreasing with
increasing optical depth.

As noted earlier, the PAH features in our pure starburst models are quite resilient even at high optical depths.
This behavior can be understood from equation~(\ref{eq:EW}): Heating by the starburst spectrum
produces dust temperature no higher than \Tin\ = 580\,K, thus the transmitted
component of the original input radiation remains significant in comparison
with the thermal emission of reprocessed radiation from the shell. As the AGN
fraction of the heating spectrum rises, PAH feature equivalent widths decrease both as a result
of their reduced prominence in the input SED and the increasing contribution of
thermal emission from the shell in the spectral vicinity of the PAH feature as a result of
its higher dust temperature.

\subsection{Spectral Slope}
\label{sec:SpectralSlope}

\ISO\ revealed that the slope of the 5--15\um\ portion of the spectrum can be used as a powerful diagnostic tool for 
identifying the presence of AGN in obscured galaxies \citep{Laurent00}. Within the general context of the unified 
model of AGN---and independent of whether the AGN is surrounded by a smooth torus or clumpy dense 
ISM clouds---dust in the proximity of the AGN is heated to near sublimation temperature ($\sim$1,000~K 
for silicate or $\sim$1,500~K for graphitic grains). The hottest surviving grains in thermal equilibrium 
with an AGN radiation field therefore emit with a peak at $\sim$3\um. Dust layers located further away from the AGN 
reprocess this radiation and are, therefore, cooler---emitting the bulk of their energy at longer mid-IR wavelengths 
\citep[e.g.,][]{Granato97, Clavel00, Murayama00, Alonso-Herrero01}. Photons from massive starbursts do not heat 
the same quantity of dust to such high temperatures and, as a result, produce much weaker 3--7\um\ dust continua. 

Infrared emission in the 15\um\ range from dust heated to $\sim$200~K can, however, be produced either by star-formation 
or by the outer layers of the clouds or shell enshrouded an AGN. This behavior of the mid-IR continuum has been successfully 
used in analysis of \Spitzer\ \IRS\ spectra to reveal the power production mechanism in local ULIRGs 
\citep[e.g.,][]{Armus07, Petric11}.
As described in \S\ref{sec:AnalysisModelsData}, in this work we choose to characterize the mid-IR slope based
upon the 14\um\ to 5.5\um\ flux density ratio since the continuum at these wavelengths is relatively free of 
strong PAH emission and silicate absorption features.

\subsection{Analysis of Models and Data}
\label{sec:AnalysisModelsData} 

Our observational sample of sources to compare to our models includes \Spitzer\ \IRS\ spectra of 103 ULIRGs from 
\citet{Armus04, Armus06, Armus07} and \citet{Desai07} which have sufficiently high signal-to-noise
ratio data to yield accurate $9.7\um$ silicate absorption feature strength measurements. 
For comparison, our sample also includes SEDs of 16 lower luminosity starburst galaxies 
from \citet{Brandl06} and 13 AGNs and quasars from \citet{Hao05} and \citet{Weedman05}. 
All spectra were extracted and calibrated using the method described in \citet{Spoon07}.

Silicate feature strengths of observed and model SEDs are
calculated from equation~(\ref{eq:Sil}) with the continuum interpolation
method described in \citet{Sirocky08}. PAH equivalent widths are calculated using the method 
described in \citet{Spoon07}. In short,
the total flux in the $6.2\um$ PAH feature is measured by integrating the flux
density above a local spline-interpolated continuum, and the equivalent width
is calculated by dividing this by the value of the interpolated 6.2\mic\
continuum. Mid-IR continuum slopes are calculated from the interpolated
5.5\mic\ flux density and the continuum flux density at 14\mic. These wavelengths
were chosen because: (1) they both fall outside the range of any significant PAH 
emission features; and (2) the latter wavelength falls between the two silicate features at 
$\sim9.7\mic$ and $\sim18\mic$.

Figure~\ref{fig:ModelParams} shows the dependence of these spectral indicators on the $5500\Angstrom$ 
optical depth, \tV, for the models shown in Figure~\ref{fig:ModelSEDs_IR}. As is evident from the figure, our 
three indicators behave quite differently for the models of pure starburst galaxies and AGN. The starburst 
SEDs have shallower silicate absorption features at high optical depth, steeper mid-IR slopes (i.e., ``cooler'' 
5.5\mic\ emission in comparison to their 14\mic\ emission), and larger PAH equivalent widths.

\section{3-D Diagnostic Diagram}
\label{sec:diagnostic}

\begin{figure*}
  \epsscale{1.18}
  \plotone{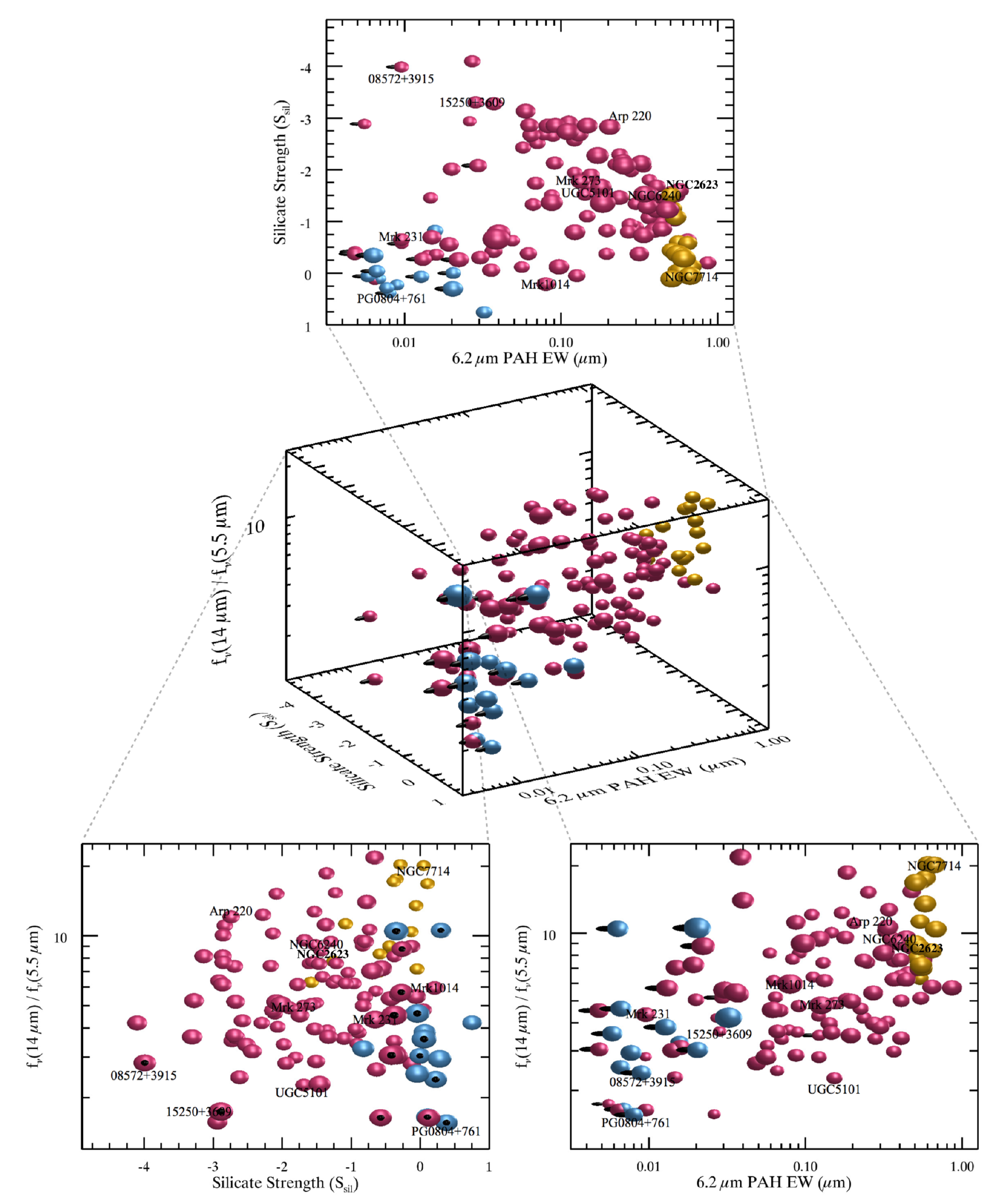}

\caption{3-D diagnostic diagram based upon the mid-IR spectral indicators described
in \S\S\ref{sec:sil}--\ref{sec:SpectralSlope} for the sample of starburst galaxies (\textit{gold spheres}), 
AGN (\textit{blue spheres}), and ULIRGs (\textit{maroon spheres}) described in
\S\ref{sec:AnalysisModelsData}. The 3-D diagnostic is presented along with its trio of 2-D
projections. The sizes of symbols indicate proximity to the viewer---i.e., a larger data point means 
that the point is closer to the observer for the given orientation of the cube. Black arrows 
(seen in face-on projections as black circles) indicate upper-limits for PAH EW. The locations of several well-known 
sources from our sample---including PG\,0804+761, our template AGN heating source, and NGC\,7714 
and NGC\,2623, our template unobscured and obscured starburst nuclear heating sources---are also indicated.}
  \label{fig:Diagnostic_Data}
  \epsscale{1.0}
\end{figure*}

\begin{figure*}
  \epsscale{1.0}
  \plotone{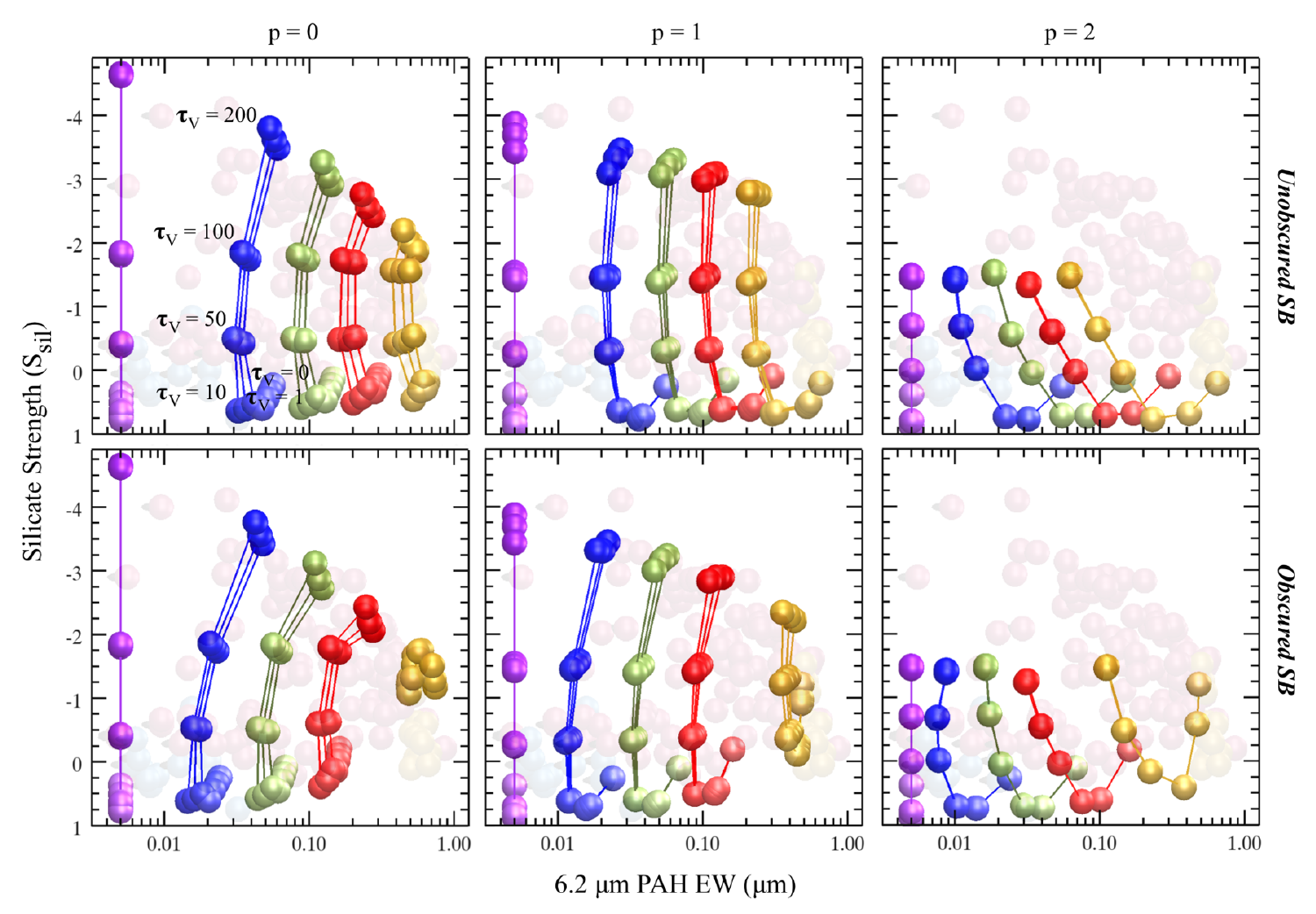}
  \plotone{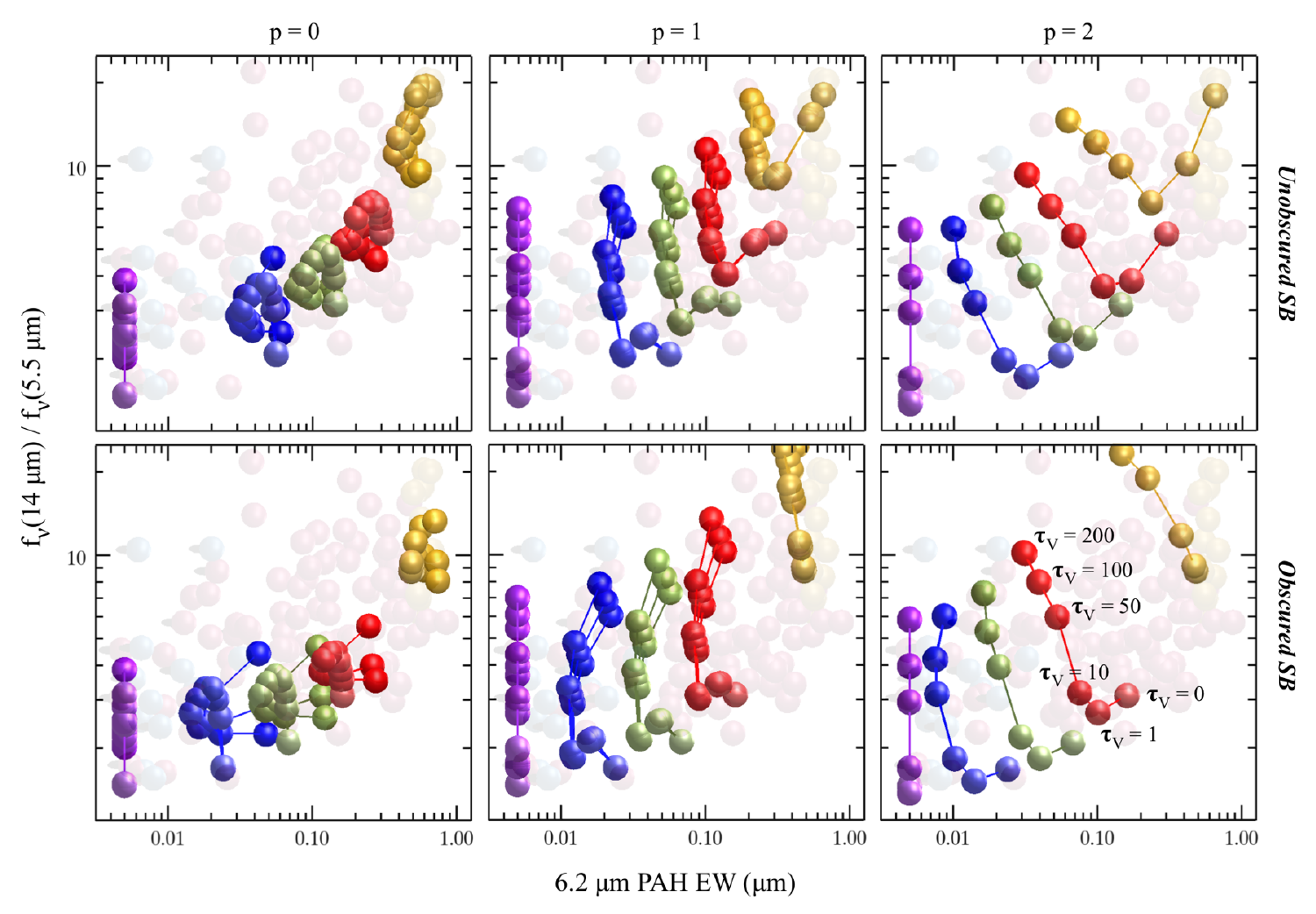}
  
\caption{Model tracks for the two-dimensional projections of silicate strength and mid-IR slope versus 
PAH equivalent width from the 3-D diagnostic diagram of Fig.~\ref{fig:Diagnostic_Data} first presented 
and discussed in \citet{Spoon07} and \citet{Laurent00}, respectively. Model tracks are 
shown for the suite of ULIRG models described in \S\ref{sec:ModelingTotalEmission} and shown in 
Fig.~\ref{fig:ModelSEDs_IR}. As in Figs.~\ref{fig:InputSEDs}--\ref{fig:ModelParams}, colors of tracks 
indicate AGN fraction, \xAGN, for that model. Models for $p = 0$, 1, and 2 are presented in the 
left, center, and right columns, respectively. Models produced using the unobscured and obscured starburst 
heating sources are shown in the top and bottom rows, respectively. Tracks are shown for $Y = 100$, 200, and 400. 
Dots along each track represent values for $\tV = 0$, 1, 10, 50, 100, and 200. Observed values for our sample 
of starburst galaxies, AGN, and ULIRGs from Fig.~\ref{fig:Diagnostic_Data} are visible beneath the tracks to 
aid in comparing the models to the data.}

  \label{fig:Diagnostics-A}
  \epsscale{1.0}
\end{figure*}

\begin{figure*}
  \epsscale{1.0}
  \plotone{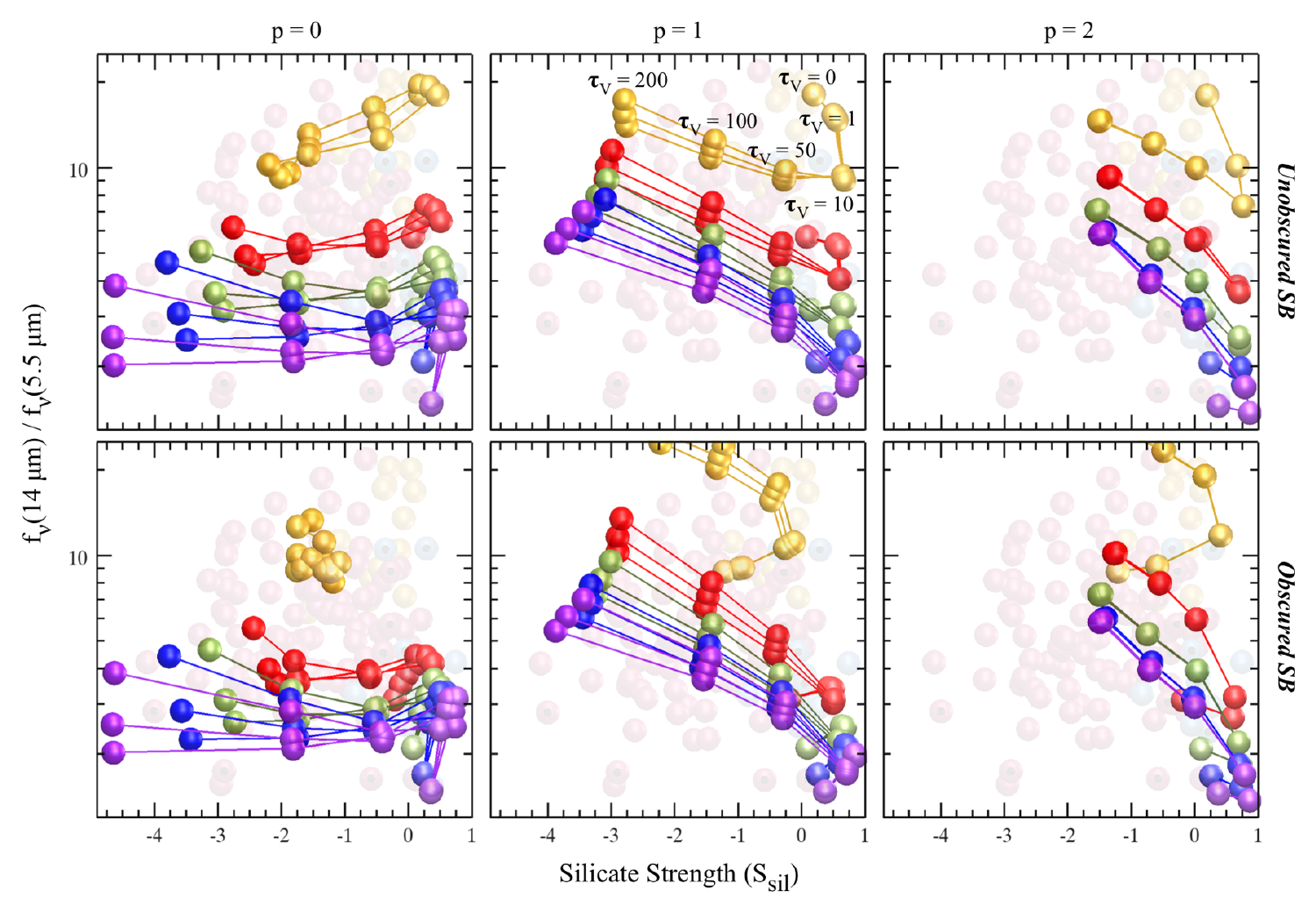}

\caption{Model tracks for the two-dimensional projection of slope versus silicate strength 
from the 3-D diagram of Fig.~\ref{fig:Diagnostic_Data}. Models are as described in 
Fig.~\ref{fig:Diagnostics-A}.}

  \label{fig:Diagnostics-B}
  \epsscale{1.0}
\end{figure*}

In Figure~\ref{fig:Diagnostic_Data}, we show a three-dimensional representation of the diagnostic features 
from \S\ref{sec:AnalysisModelsData} for the archival \Spitzer\ \IRS~data of our sample. Along with the
3-D view, we also show its trio of 2-D projections. Two of these projections have been discussed extensively
before, with $\Sil$ versus $6.2\um$ PAH EW and mid-IR spectral slope versus $6.2\um$ PAH EW having 
first been presented in \citet{Spoon07} and \citet{Laurent00}, respectively. We suggest that these two diagrams 
are best thought of as two-dimensional projections of a more fundamental three-dimensional phase space of
mid-IR ULIRG properties. Sources from the AGN and starburst subsets of our sample tend to live in the 
same region of this phase space, while ULIRGs (due to their diverse nature) tend to maximally fill the space. 
The locations of several well-known sources from our sample---including our template AGN and starburst 
heating sources---are shown on each projection.

Figures~\ref{fig:Diagnostics-A} and \ref{fig:Diagnostics-B} show the trio of 2-D projections from
the 3-D diagnostic diagram in Figure~\ref{fig:Diagnostic_Data} for the suite of ULIRG models 
described in \S\ref{sec:ModelingTotalEmission} and presented in Figure~\ref{fig:ModelSEDs_IR}. 
These models assume $\xShell = 1$ (i.e., a fully covered central source; see eq.~[\ref{eq:f_nucleus}]), 
$\xPAH = 0.1$ (i.e., PAHs contributing 10\% of the total luminosity of disk star formation; see 
eq.~[\ref{eq:f_disk}]), and $\alpha_\mathrm{disk} = 0.1$ (i.e., a host galaxy disk contribution of 
10\% the luminosity of the nuclear starburst; see eq.~[\ref{eq:L_disk}]). Overall, the models 
do an excellent job at reproducing the location of the ULIRGs (seen projected faintly behind each model 
diagram) in this three-dimensional phase space of mid-IR spectral properties.  By mixing pure 
AGN and starburst models, we succeed in mapping out the entire space occupied by the ULIRGs, 
as well as those of the known pure AGN and starburst samples.

Of particular interest is the result that the somewhat paradoxical subclass of ULIRGs which 
simultaneously display extremely deep silicate features ($\Sil < -4$) and warm mid-IR colors 
($F_\nu(14\um) / F_\nu(5.5\um) < 5$) can \textit{only} be reproduced by AGN-dominated models. 
This implies that these galaxies must have a buried AGN contributing significantly to their total power. 
Starbursts that contribute more than $\sim25\%$ of the total power always generate mid-IR spectra 
with less silicate absorption and steeper slopes. Furthermore, in cases where degeneracies as to the 
nature of a ULIRG exist in one or even two projections, our new interpretation of the trio of spectral 
indicators as a 3-D diagnostic phase space can sometimes yield a diagnoses by adding input from 
the third projection.

Another perhaps less satisfying result is that in some regions of phase
space---e.g., much of the spectral slope versus silicate strength projection---the
models with different values of $p$ are degenerate. While the extremes or edges 
of the space are only populated by specific models, there are large regions where real 
galaxies live that can be accommodated by a wide range of model parameters.  Determining 
the exact radial thickness and density distribution of the obscuring medium for
many ULIRGs is difficult in the context of these models, especially when mixing
starburst and AGN power sources in composite systems.  The models are nonetheless 
useful as they, at a minimum, allow us to use simple observables to place important 
constraints on the power sources and the obscuring medium around ULIRG nuclei.

\section{Discussion}
\label{sec:discussion}

\begin{figure*}
  \epsscale{1.1}
  \plotone{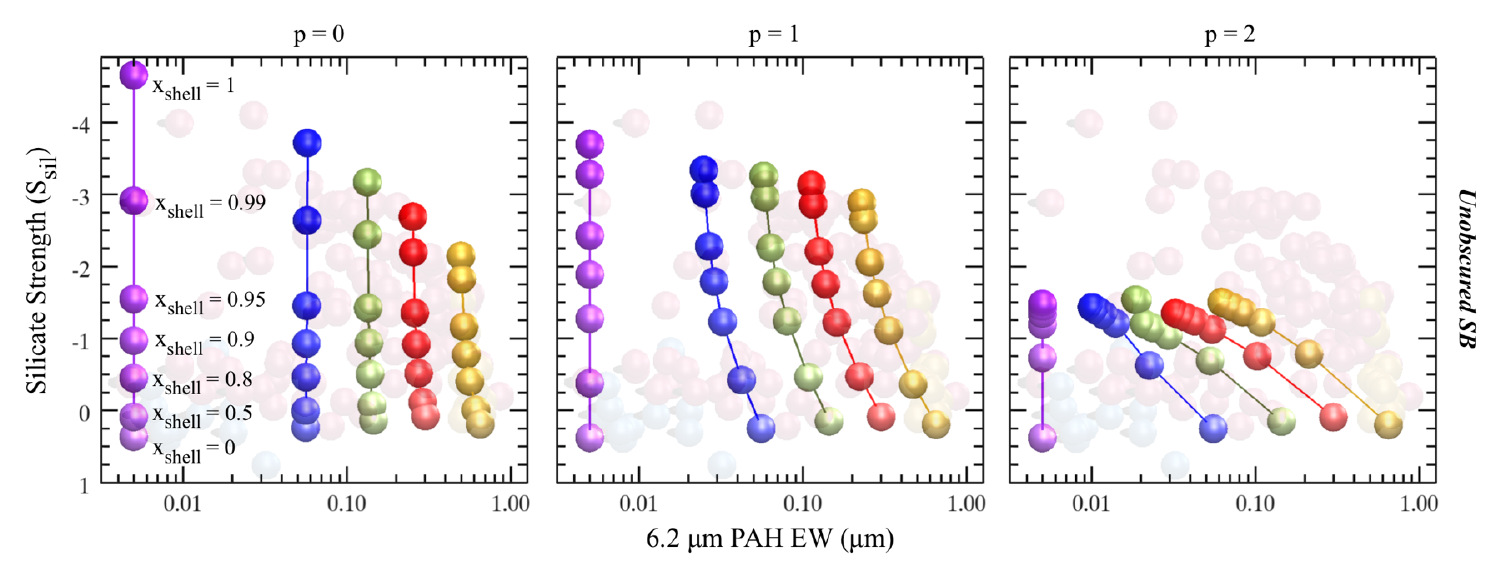}
  
  \caption{Model tracks for the case in which the obscuring dust shell does not fully
  cover the AGN and/or starburst ULIRG central engine. The top dot along each track
  matches the top dots of the $Y = 200$, $\tV = 200$, unobscured starburst models of
  Fig.~\ref{fig:Diagnostics-A} for the indicated values of $p$. Each subsequent
  dot along the track represents a change in the covering factor from fully covered at the
  top ($\xShell = 1$) to completely uncovered ($\xShell = 0$) at the bottom. Note the
  rapidity with which the tracks cross the ``gap'' in the observed sources visible beneath
  the model tracks. Even a relatively small ($\sim 5$--10\%) ``keyhole'' view through the 
  obscuring medium is sufficient to make the source appear nearly unobscured in the mid-IR.}
  \label{fig:Clumpy_Models}
  \epsscale{1.0}
\end{figure*}

In Figure~\ref{fig:Clumpy_Models}, we explore the effects of a clumpy obscuring medium
surrounding the central source. Here the model provides a ``keyhole'' view through to
a fraction of the underlying AGN and/or starburst-powered nuclear source. Of particular note, 
when only $\sim$10\% of the underlying source is revealed (i.e., a 90\% covering fraction), 
the highly obscured AGN located towards the top left of the silicate strength versus $6.2\um$
PAH EW plot quickly move from the deeply obscured region to the apparently unobscured 
region at the bottom of the diagram. As such, only completely obscured AGN with covering 
factors near unity can be found towards the upper left of this diagram (since even a 5\%
``keyhole'' view is sufficient to move a source towards the region occupied by unobscured AGN).

This may explain why we find so few moderate-to-deeply buried ULIRGs with low PAH equivalent 
widths: ULIRGs with low PAH equivalent width either tend to have little-to-no apparent 
obscuration or they appear to be deeply buried (with very few in-between). 
Our models suggest that this change in the silicate strength is very sensitive to small 
changes in the fraction of the central source that is visible. As such, the removal of the obscuring 
dust from a highly extinguished, low PAH equivalent width source may or may not happen quickly.
But as soon as a small window to the central source is revealed, the silicate strength changes 
dramatically. The dearth of real sources \citep[see][]{Spoon07} and model points 
(see Fig.~\ref{fig:Diagnostic_Data}) with weak PAH emission at intermediate-to-large silicate depths 
may therefore be a result of the fact that the spectral properties of galaxies change easily and 
rather dramatically in this projected plane (see also \citealt{2009MNRAS.399..615R} for an alternate 
theoretical interpretation of the ``Spoon'' diagram). We note that this sensitivity to the geometry of the 
obscuring medium means that the distribution of sources in this plane should not necessarily be 
used to infer the relative amount of time a galaxy spends in the obscured and unobscured states.

\begin{figure}
  \epsscale{1.01}
  \plotone{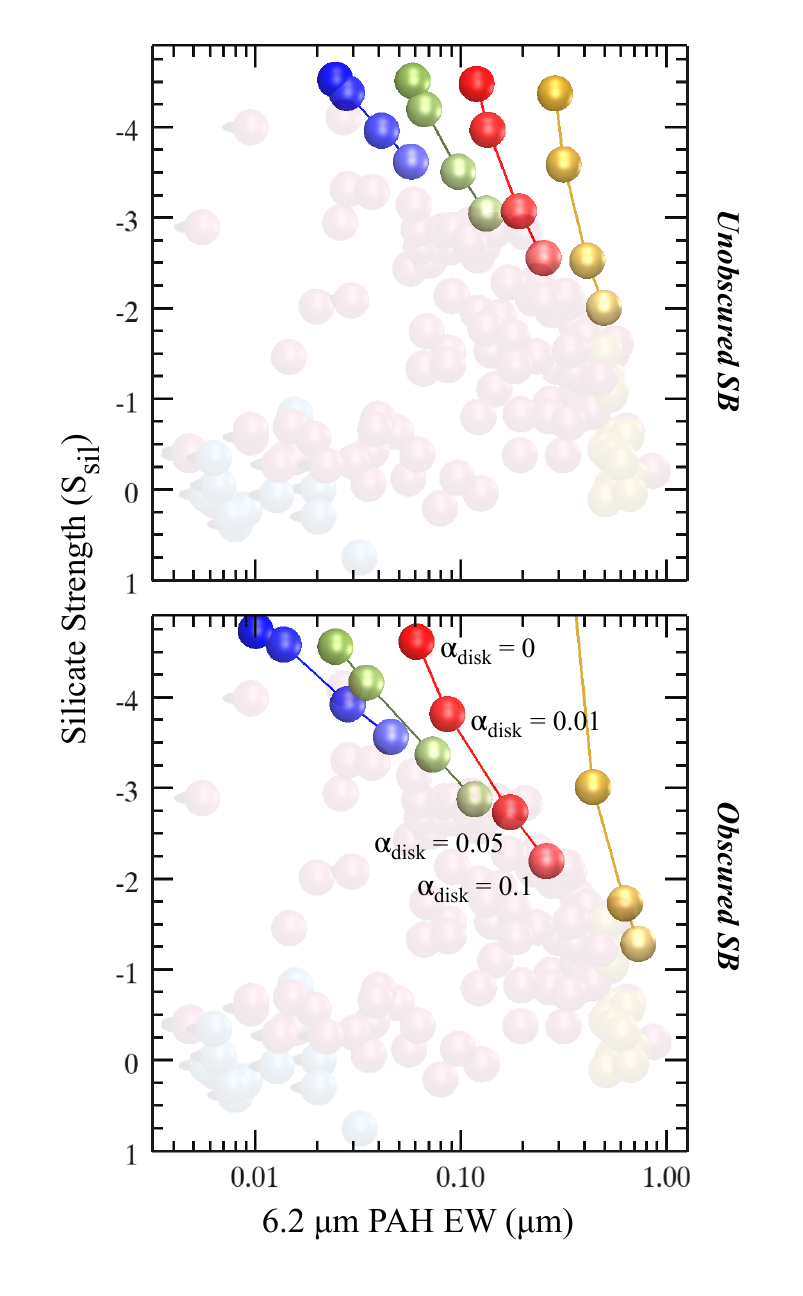}

\caption{The effect of adding different amounts of host galaxy disk emission (dominated by unobscured 
PAHs) to our most deeply buried models. The top point on each track does not include any added
disk emission (i.e., $\alpha_\mathrm{disk} = 0$). Each subsequent point from top-to-bottom is obtained by 
adding disk emission at a level of 1\%, 5\%, and 10\% of the nuclear starburst luminosity 
(i.e., $\alpha_\mathrm{disk} = 0.01$, 0.05, and 0.1) to the $p = 0$, $Y = 200$, and $\tV = 200$ model 
for each value of \xAGN\ (we omit the pure AGN case since it has no nuclear starburst luminosity and 
thus no disk component). As is evident, even this relatively small amount of unobscured PAH emission 
is sufficient to move starburst galaxies from unoccupied regions of the diagram into regions that are 
populated by the observed sources visible beneath the model tracks.}

  \label{fig:Models+Disk}
  \epsscale{1.0}
\end{figure}

In Figure~\ref{fig:Models+Disk}, we show the effect of adding small amounts of
unobscured PAH emission from the host galaxy disk to the models. Specifically, 
we use values of $\alpha_\mathrm{disk} = 0$, $0.01$, $0.05$, and $0.1$
(see eq.~[\ref{eq:L_disk}]), along with $\xPAH = 0.1$  (see eq.~[\ref{eq:f_disk}])
to model the effect of a contribution from unobscured host galaxy disk emission at the 
0\%, 1\%, 5\%, and 10\% level relative to the bolometric nuclear star-formation luminosity.  
Even these small amounts of PAH emission are sufficient to move galaxies from the upper-left to
the lower-right in the diagram. This naturally explains why highly obscured starburst dominated 
ULIRGs are rarely seen.  These sources may exist, but even a relatively small amount of disk 
PAH emission rapidly makes them appear much less obscured and moves them away from 
the upper part of Figure~\ref{fig:Diagnostics-A}. Just as importantly, this also implies that ULIRGs 
with deep silicate absorption and low PAH equivalent width cannot have more than $\sim$10\% 
of their total emission coming from unobscured star-formation. Otherwise, they would move 
towards the pure starburst galaxies in the lower-right of the diagram.

The vast majority of low-redshift ULIRGs were formed from major mergers of
gas-rich disk galaxies. These disks will always have a small amount of residual
star-formation that, although dwarfed by the subsequent central starburst,
likely contributes a small amount of PAH emission to the
mid-IR spectrum---even when that spectrum is centered on the nucleus. The \IRS\ slits (which
match the \Spitzer\ beam) are relatively large in projection ($3.6^{\prime\prime}$ and $11^{\prime\prime}$ 
for SL and LL, respectively), so disk emission from the surrounding host galaxy is 
likely to leak into the ``nuclear'' spectrum. In the context of our model, 
it is the relative amount of central starburst and disk PAH emission
that controls where ULIRGs fall in Figure~\ref{fig:Models+Disk}---even a small amount of 
circumnuclear star formation makes them appear much less obscured. 

At high-redshift where apertures cover many kiloparsecs, it is likely to be extremely difficult 
to find highly-obscured systems in which even a small amount of disk star-formation 
is not present (whether or not a system has been involved in a major merger). This may be 
consistent with the findings of \citet{Pope08} and \citet{Delmestre09} who, 
on average, find stronger PAH and shallower silicate absorption in $z\approx2$ SMGs 
compared to local ULIRGs and interpret this as extended, less obscured star formation. 

\begin{figure*}
  \epsscale{1.0}
  \plotone{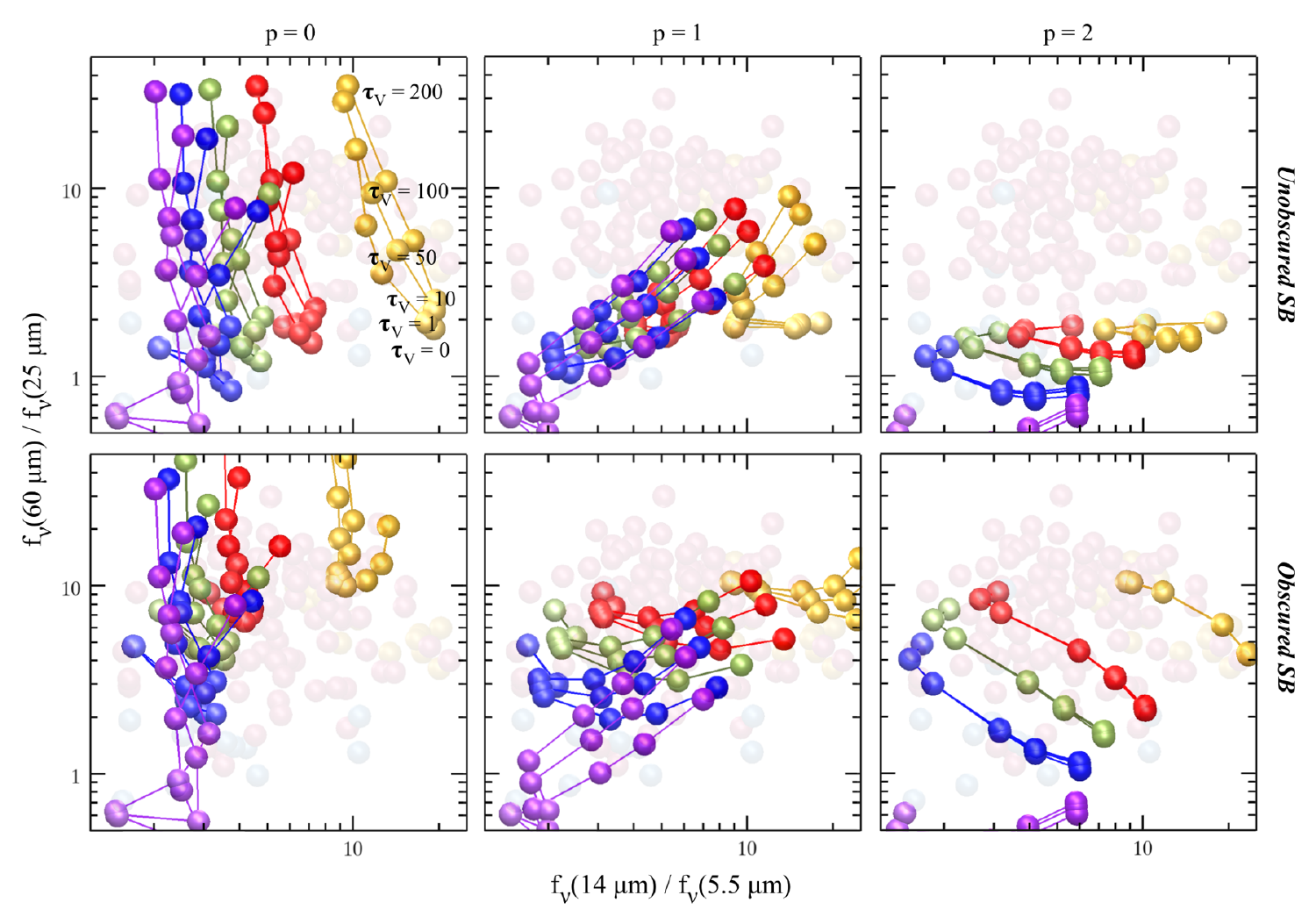}

\caption{Model tracks along a plot of far-IR to mid-IR slope. The models are as described in 
Fig.~\ref{fig:Diagnostics-A}, and span the full range of colors of our observational sample  
(visible beneath the tracks) described in \S\ref{sec:AnalysisModelsData}. Note that the upper
regions of this color-color phase space (associated with ``cool'' far-IR colors) can only be 
reached by models having a constant density ($p = 0$) obscuring shell. Density profiles
which fall-off faster with radius do not have sufficient dust at large radii to produce enough far-IR
emission to populate these regions. For $p = 1$, the model tracks are relatively insensitive to the
choice of starburst template, but the obscured starburst models with $p = 0$ and $p = 2$ are 
cooler than the unobscured models for a given $\tV$.}

  \label{fig:Diagnostic_FIR-vs-MIR}
  \epsscale{1.0}
\end{figure*}

In Figure~\ref{fig:Diagnostic_FIR-vs-MIR}, we present the far-IR versus mid-IR colors 
of our models overlaid atop the positions of the sources in our archival sample. While we have not 
developed a comprehensive model of the far-IR emission emerging from ULIRG nuclei 
and their host galaxy disks, our suite of models nonetheless does a good job at fitting these far-IR 
colors. This result provides additional evidence that the assumptions used in generating the models 
for the ULIRG nuclei and host galaxies are reasonable.

According to our models, the obscuring shells surrounding the nuclei of sources with ``cool'' 
far-IR colors (i.e., $f_\nu(60\um) / f_\nu(25\um) \ga 10$) in the upper regions of the 
color-color phase space of Figure~\ref{fig:Diagnostic_FIR-vs-MIR} must have relatively flat ($p = 0$)
mass density profiles. The densities of the $p = 1$ and $p = 2$ models fall-off faster with radius 
and thus contain insufficient dust at large radii to produce the quantity of far-IR emission needed to populate 
the ``cool'' regions of this phase space. Note also that for the $p = 1$ case, the model tracks are relatively 
insensitive to the choice of starburst template (i.e., unobscured versus obscured); however, the obscured 
starburst models with $p = 0$ (and to a lesser extent $p = 2$) are clearly cooler than the unobscured
models for a given value of $\tV$.

Finally, the superior sensitivity and unprecedented angular resolution of \JWST\ 
will soon provide a unique opportunity to use mid-IR spectroscopy to dissect the dust enshrouded 
central regions of both nearby nuclei and more distant galaxies. These observations will benefit from 
a more than ten-fold improvement in angular resolution compared to \Spitzer\ \IRS\ for the $\sim$3000 
local galaxies detected by that observatory \citep[see][]{2016MNRAS.455.1796H}. \JWST\ will
also allow us to explore many more than the $\sim$300 moderate to high-z galaxies for which \Spitzer\
has provided low-resolution spectra \citep[see][]{2015ApJ...814....9K}.

The integral field units (IFUs) of the \JWST\ mid-infrared instrument (MIRI) will enable the same 
kind of analysis using the well known emission features in the 5--28$\um$ range that we have 
performed in this work with \IRS\ spectra, but for numerous distinct resolved areas of target galaxies. 
Of particular interest will be the ability of the IFUs of the near infrared spectrograph (NIRSpec) on \JWST\ 
to probe the spatial distribution of thermal dust emission near an AGN as well as the circumnuclear 
$3.3\um$ PAH feature at redshifts up to 0.7 (and thus spatial scales of $\la700$\,pc). Such analyses of 
the detailed spectral features will provide additional constraints to quantify the AGN/star-formation power 
in nearby and distant galaxies currently limited to exploration via broad-band imaging alone 
\citep[see][]{2017arXiv170609056K}.

\section{Summary}
\label{sec:Summary}

We have presented a suite of radiative transfer models of the SEDs of deeply
buried star-formation and/or AGN-powered ULIRGs and have compared them to the
SEDs of a large sample of $\sim$100 ULIRG spectra taken with the \IRS\ on
\Spitzer. We find that the properties that dominate the observed mid-IR spectra
of ULIRGs (specifically their mid-IR colors, $6.2\um$ PAH equivalent widths, and
$9.7\um$ silicate feature strengths) are consistent with the range of values
obtained from our radiative transfer models of a spherically symmetric shell of 
obscuring dust surrounding a star-formation and/or AGN powered central source.

We find that the commonly used PAH equivalent width versus silicate strength and
PAH equivalent width versus mid-IR slope diagrams are naturally described as
two-dimensional projections of a more fundamental 3-D phase space
of silicate strength, PAH equivalent width, and mid-IR spectral slope. 

In addition to demonstrating the general consistency of our simple ULIRG models
with observational data across this 3-D phase space (and even into the far-IR), 
we have shown:

\begin{enumerate}

\item Our models agree with observations that the most heavily obscured
    ULIRGs (i.e., $S_{\rm Sil} \la -4$) have flat mid-IR slopes and an
    apparent ``excess'' of hot dust emission (i.e., $f_\nu(14.5\um) /
    f_\nu(5.5\um) < 5$). Furthermore, our models show that at least 75\% of
    the bolometric luminosities of these sources must be powered by an AGN,
    with the most deeply buried likely being 100\% AGN-powered. The bolometric
    luminosities of slightly less obscured ULIRGs (i.e., $-3 \la S_{\rm Sil}
    \la -2.5$) may be dominated by either an obscured AGN or an obscured
    starburst. This degeneracy can be (at least partially) broken using the
    mid-IR slope.

\item Transitioning from a deeply buried (i.e., $S_{\rm Sil} < -3$) and low
    PAH equivalent width (< 0.01) AGN-dominated source to a relatively
    unextinguished ($S_{\rm Sil} > -1$) AGN-powered SED can be accomplished
    by unveiling just 5--10\% of the underlying nuclear power source. In other words,
    only a small window through a clumpy dust shell is required for a source
    to ``jump'' from the upper to lower branch of the PAH equivalent width
    versus silicate strength diagram. This may explain the rarity of sources
    detected with \Spitzer\ \IRS\ having low PAH equivalent width and -3 <
    $S_{\rm Sil} < -1$. And it suggests that the distribution of sources in this plane 
    should not necessarily be used to infer the relative amount of time a galaxy spends 
    in the obscured and unobscured states.

\item The absence of sources with observed SEDs that display both deep
    silicate absorption and strong PAH emission features is naturally
    explained by the presence of a small amount of unobscured star-formation
    (and therefore PAH emission) from the host galaxy disk. Unobscured star
    formation with a total PAH luminosity of only $\sim$1\% that of the
    buried starburst (equivalent to a total disk luminosity of $\sim$10\%
    the nuclear star-forming luminosity in our model) is sufficient to drive 
    sources to the lower right of the PAH equivalent width versus silicate 
    strength plot. On a related note, as shown in Figure~\ref{fig:Models+Disk}, 
    sources with $S_{\rm Sil} \la -2$ cannot 
    have more than 10\% of their bolometric emission coming from unobscured 
    star-formation. If they did, their deep silicate features would be partially
    ``filled-in,'' pushing them off the upper-branch of the diagram down
    towards the location populated by unobscured starburst galaxies.

\end{enumerate}

\acknowledgements 
The authors greatly appreciate the detailed and constructive comments of the anonymous referee which 
substantially improved the clarity of the manuscript. 

This work is based [in part] on observations made with 
the \LongSpitzer, which is operated by the Jet Propulsion Laboratory, California Institute of Technology 
under NASA contract 1407.  Support for this work was provided by NASA through Contract Number 
1257184 issued by JPL/Caltech. M.E. acknowledges NASA support through grants NNX11AD76G and 
NNX15AC83G. This research has made use of the NASA/IPAC Extragalactic Database (NED) which is 
operated by the Jet Propulsion Laboratory, California Institute of Technology, under contract with the 
National Aeronautics and Space Administration. T.D-S. acknowledges support from ALMA-CONICYT 
project 31130005 and FONDECYT regular project 1151239.

\bibliographystyle{apj}


\end{document}